\documentclass[reqno,11pt]{amsart}
\usepackage{fullpage}
\usepackage{graphics,amsmath,amssymb,amsthm,accents,epic}
\usepackage{mathrsfs}
\usepackage{cite}
\usepackage[top=1 in,bottom=1 in,left=1.0 in,right=1.0 in]{geometry}
\usepackage{bm}
\usepackage{subfigure}
\usepackage{latexsym}
\usepackage{xcolor}
\usepackage[colorlinks,linkcolor=black,anchorcolor=blue,citecolor=blue]{hyperref}
\usepackage{amsfonts}
\usepackage{mathtools}
\usepackage{physics}
\usepackage{multirow}
\usepackage{caption}
\usepackage{ulem}

\allowdisplaybreaks
\numberwithin{equation}{section}

\newtheorem{prop}{Proposition}[section]
\newtheorem{theorem}{Theorem}[section]

\newtheorem{remark}{Remark}[section]

\newcommand{\wh}{\widehat}
\newcommand{\wt}{\widetilde}

\newcommand{\bM}{\boldsymbol{M}}
\newcommand{\bL}{\boldsymbol{L}}
\newcommand{\br}{\boldsymbol{r}}

\newcommand{\ba}{\boldsymbol{a}}
\newcommand{\bC}{\boldsymbol{C}}
\newcommand{\bK}{\boldsymbol{K}}
\newcommand{\bS}{\boldsymbol{S}}

\newcommand{\bV}{\boldsymbol{V}}
\newcommand{\bW}{\boldsymbol{W}}
\newcommand{\bX}{\boldsymbol{X}}
\newcommand{\bY}{\boldsymbol{Y}}
\newcommand{\bZ}{\boldsymbol{Z}}
\newcommand{\bu}{\boldsymbol{u}}
\newcommand{\bI}{\boldsymbol{I}}
\newcommand{\bv}{\boldsymbol{v}}
\newcommand{\bw}{\boldsymbol{w}}
\newcommand{\bG}{\boldsymbol{G}}

\newcommand{\be}{\boldsymbol{e}}
\newcommand{\bH}{\boldsymbol{H}}
\newcommand{\bJ}{\boldsymbol{J}}
\newcommand{\bF}{\boldsymbol{F}}

\newcommand{\bs}{\boldsymbol{s}}
\newcommand{\bh}{\boldsymbol{h}}
\newcommand{\bA}{\boldsymbol{A}}
\newcommand{\bB}{\boldsymbol{B}}
\newcommand{\bt}{\boldsymbol{t}}
\newcommand{\mbx}{\boldsymbol{x}}
\newcommand{\mby}{\boldsymbol{y}}

\newcommand{\nn}{\nonumber}
\newcommand{\st}{\hbox{\tiny\it{T}}}

\begin{document}

\title{Discrete four-component massive Thirring model}	

\author{Xiao-gang Mu,~~Jing Wang,~~Song-lin Zhao$^{*}$\\
\\\lowercase{\scshape{
School of Mathematical Sciences, Zhejiang University of Technology, Hangzhou 310023, P.R. China}}}
\email{Corresponding Author: songlinzhao@zjut.edu.cn}

\begin{abstract}

We construct a discrete four-component massive Thirring model (MTM) within the Cauchy matrix framework. 
A connection between the dependent variables of the discrete four-component MTM and those of the discrete Mikhailov model is established.
Taking two successive continuum limits, we first reduce this discrete four-component MTM to its semi-discrete counterpart, 
and further arrive at the continuous four-component MTM.
Solving the determining equation set enables us to construct AKNS-type and KP-type solutions, including explicit soliton solutions and multiple-pole solutions.
Imposing the local reduction and nonlocal reduction on the discrete four-component MTM yields the discrete MTM and the reverse-$(n,m)$ discrete MTM, respectively.
For these two reduced systems, we prove that the former admits AKNS-type solutions, whereas the latter possesses both AKNS-type and KP-type solutions.
 
\end{abstract}


\maketitle

\section{Introduction}

The massive Thirring model (MTM) has its origins in Thirring's 1958 study of an exactly solvable massless fermion theory with a
current--current interaction in $(1+1)$ dimensions \cite{T1958}. In 1975, Coleman established the quantum correspondence between the
MTM and the sine-Gordon (sG) model, identifying the sG soliton with the fundamental Thirring fermion \cite{C1975}. Mikhailov
investigated the integrability of the classical MTM in 1976 \cite{M1976}. In light-cone coordinates, the classical model
takes the form
\begin{subequations}\label{MTM-1}
    \begin{align}
        \mathrm{i}\chi_{1,x}-\frac{m}{2}\chi_2
        -g|\chi_2|^2\chi_1 &= 0, \label{MTM-1a}\\
        \mathrm{i}\chi_{2,t}-\frac{m}{2}\chi_1
        -g|\chi_1|^2\chi_2 &= 0. \label{MTM-1b}
    \end{align}
\end{subequations}
Here $\chi_1$ and $\chi_2$ are the two chiral components of the fermion field $\chi=(\chi_1,\chi_2)$, $m$ is the mass parameter,
and $g$ is the self-interaction coupling constant. The notation $|\chi_j|^2=\chi_j\chi_j^*$ denotes the squared modulus, with
${}^*$ indicating complex conjugation. Kuznetsov and Mikhailov developed the inverse-scattering formulation
of the classical MTM and established its complete integrability \cite{KM1977}. Kaup and Newell examined its classical connection with
the sG equation \cite{KN1977}, and Morris exhibited a prolongation structure associated with an $\mathrm{SL}(2,\mathbb{R})$
connection \cite{M1979}. Subsequent studies clarified the relation between the MTM and derivative nonlinear Schr\"odinger equations
through quadratic spectral bundles \cite{GIK1980} and linearization methods \cite{NCQL1983}. The direct linearization approach was also
applied to multicomponent generalizations of the MTM \cite{LCN1989}. Skrypnyk later constructed integrable generalizations of the MTM
using graded loop algebras and classical $R$-operators \cite{S2008}. Franca, Gomes, and Zimerman showed that the first negative flow
of the Kaup--Newell hierarchy yields the Mikhailov model and demonstrated its equivalence to the MTM \cite{FGZ2013}.

Various explicit solutions of the MTM have been constructed, covering Bäcklund transformations in the geometric Lax framework \cite{P1981}, 
multi-soliton solutions via system reduction \cite{D1979,DHS1984}, and algebro-geometric quasi-periodic, finite-band and finite-gap 
solutions \cite{D1982,D1978,PG1979,B1985,W1993,EGH2000,EEH2003}, symmetry-preserving polynomial 
Darboux matrices \cite{D2015}, $\tau$-function descriptions with bright/dark solitons and breathers \cite{CF2023}, as well as 
higher-order and general rogue-wave solutions \cite{GWCH2017,CYF2023}.

In \cite{BG1987}, a four-component version of the MTM was proposed
\begin{subequations}
\label{eq:continuous-mtm}
	\begin{align}
		\mathrm{i}q_{1,t}+q_3+q_1q_3q_4&=0,\label{eq:continuous-mtm-a}\\
		\mathrm{i}q_{2,t}-q_4-q_2q_3q_4&=0,\label{eq:continuous-mtm-b}\\
		\mathrm{i}q_{3,x}+q_1+q_1q_2q_3&=0,\label{eq:continuous-mtm-c}\\
		\mathrm{i}q_{4,x}-q_2-q_1q_2q_4&=0,\label{eq:continuous-mtm-d}
	\end{align}
\end{subequations}
where $q_j=q_j(x,t),~(j=1,2,3,4)$ are independent complex-valued scalar fields.
The system \eqref{eq:continuous-mtm} is equipped with a Lax pair, the conservation law $(q_1q_2)_t+(q_3q_4)_x=0$,
and exhibits a ``residual'' gauge invariance; it is also manifestly Lagrangian and 
admits explicit $N$-soliton solutions with nonvanishing constant backgrounds \cite{BGK1993,BG1993}. 
Tsuchida and Wadati generalized the system \eqref{eq:continuous-mtm} into a matrix form, whose Lax pair was shown explicitly \cite{TW1999}.

The system \eqref{eq:continuous-mtm} admits the local reduction $(q_2,q_4)=(q_1^*,q_3^*)$ giving
\begin{subequations}
\label{eq:local-mtm}
\begin{align}
	&\mathrm{i}q_{1,t}+q_3+q_1|q_3|^2=0,\\
	&\mathrm{i}q_{3,x}+q_1+q_3|q_1|^2=0, 
\end{align}
\end{subequations}
which can be recovered from the two-field MTM displayed at the beginning by
setting $\chi_1=q_3$ and $\chi_2=q_1$ with $m=-2,~g=-1$. Furthermore, we find that system \eqref{eq:continuous-mtm} possesses two different nonlocal reductions. 
\begin{itemize}
\item
Nonlocal reduction-I: $(q_2,q_4)=(q_{1,-},q_{3,-})$\footnote{For the function $f:=f(x_1,x_2)$, notation $f_{-}$ means $f_{-}=f(-x_1,-x_2)$.
When both the independent variables $x_1$ and $x_2$ are discrete, it is necessary to figure out
$\wt{f}_{-}=f(-x_1-1,-x_2)$, $\wh{f}_{-}=f(-x_1,-x_2-1)$
and $\wh{\wt{f}}_{-}=f(-x_1-1,-x_2-1)$, respectively.},
\begin{subequations}
\label{eq:nonlocal-mtm-1}
	\begin{align}
		& \mathrm{i}q_{1,t}+q_3+q_1q_3q_{3,-}=0,\\
		& \mathrm{i}q_{3,x}+q_1+q_1q_3q_{1,-}=0.
	\end{align}		
\end{subequations}
\item
Nonlocal reduction-II: $(q_2, q_4)=(-\mathrm{i}q_{1,-}^*, \mathrm{i}q_{3,-}^*)$,
\begin{subequations}\label{eq:nonlocal-mtm}
\begin{align}
& \mathrm{i}q_{1,t}+q_3+\mathrm{i}q_1q_3q_{3,-}^*=0, \\
& \mathrm{i}q_{3,x}+q_1-\mathrm{i}q_3q_1q_{1,-}^*=0.
\end{align}
\end{subequations}
\end{itemize}
We refer to both systems \eqref{eq:nonlocal-mtm-1} and \eqref{eq:nonlocal-mtm} as nonlocal MTM systems.
Very recently, Wang et al. utilized bilinearization reduction technique to investigate solutions of the nonlocal MTM \eqref{eq:nonlocal-mtm-1} \cite{WLWZ2026}. 
Moreover, relation between the nonlocal MTM and the nonlocal Fokas-Lenells equation was also discussed. On the nonlocal integrable systems,
one can refer to a series of papers \cite{AbMu-2013,Fokas-2016,AM-2016,AM-2017,Yang,Lou,Zhou,Chen,LWZ,XZS,FZF2026}. 

There exist many works related to discretization of the two-/four-component MTM in light-cone coordinates. 
The first result on the integrable discretizations of the MTM system in light-cone coordinates goes back to the work of Nijhoff et al. in \cite{NCQ1983}. 
A semi-discrete route based on discrete derivative nonlinear Schr\"{o}dinger hierarchies was developed by Tsuchida, 
covering Thirring-type systems with conjugation reductions \cite{T2002} and a light-cone semi-discrete MTM from the negative flow of 
the discrete Chen--Lee--Liu hierarchy \cite{T2015}. Recent fully discrete results include Tsuchida's zero-curvature construction 
with matrix formulation, conjugation reduction, Darboux transformation and Yang-Baxter map \cite{T2024}, and Chen and Feng's
bilinear KP-hierarchy reduction with multi-bright-soliton solutions \cite{CF2025}. For the discretization of the MTM
in laboratory coordinates, one can refer to two papers  \cite{JP2019,CP2019}. Very recently, Tsuchida proposed an integrable 
semi-discrete MTM equation in non-characteristic coordinates (intermediate between light-cone and laboratory coordinates), 
and its Lax pair representation associated with soliton solutions has been investigated \cite{T2025}.

The Cauchy matrix approach has emerged as a powerful algebraic framework for investigating integrable systems and constructing their exact solutions. 
Originally developed in the study of discrete integrable equations \cite{NAH2009,ZZ2013}, this approach exploits the algebraic structure of Cauchy matrices and 
Sylvester-type matrix equations to establish connections between spectral parameters, nonlinear evolution equations, and solutions beyond solitons \cite{XZZ2014}.
Subsequent developments have extended its applicability to continuous, semi-discrete and discrete integrable equations, including elliptic Adler--Bobenko--Suris
lattice equations \cite{NA2010}, lattice Boussinesq-type equations \cite{ZZS2012,FZZ2012}, Kadomtsev--Petviashvili (KP) type equations \cite{FZ2013,FZ2022}, 
Ablowitz--Kaup--Newell--Segur (AKNS) systems \cite{Z2016,Z2018,HSZ2025}, 
self-dual Yang--Mills equations \cite{LQYZ2022,LQZ2023}, modified Korteweg--de Vries-sG (mKdV-sG) equation \cite{CWZ2023},
semi-discrete Toda and sG systems \cite{SLZZZ2025}, as well as non-isospectral equations \cite{TLZ2022,TLLLY}.
A major advantage of this method is that it provides a unified algebraic framework for constructing multisoliton solutions,
deriving Lax pairs and $\tau$-functions, and exploring the structural relationships between different integrable models. More recently, the approach has been 
further developed for the systematic construction of integrable hierarchies, including the isospectral KdV hierarchy \cite{HLZ2026,HZ2026} and 
non-isospectral KdV hierarchy \cite{TLLLY2026}, highlighting its potential for studying higher-order integrable equations and their exact solutions. 

In \cite{ZMZ2026,MZ2026}, we used the Cauchy matrix approach to construct integrable discretizations of the Mikhailov model, 
the Gerdjikov--Ivanov model, and its higher-order generalization. In the present work, we aim to construct a discrete counterpart of 
the four-component MTM \eqref{eq:continuous-mtm} via the Cauchy matrix approach and investigate its local and nonlocal conjugation reductions. 
In our formulation, a general set of determining equations (DES) associated with the KP and AKNS schemes is introduced to define the master functions $\bS^{(i,j)}$. 
The entries of  $\bS^{(0,0)}$, $\bS^{(-1,0)}$ and $\bS^{(-1,-1)}$ are used to specify the dependent variables of the discrete four-component MTM. 
A connection between the discrete four-component MTM and the discrete Mikhailov model is shown.
We further examine successive formal continuum limits that link this discrete system to a semi-discrete four-component
MTM and subsequently to the continuous four-component MTM. 
By solving the DES, explicit solutions of the resulting discrete four-component MTM, including solitons and multiple-pole solutions, are derived in 
both AKNS-type and KP-type realizations. 
For the obtained discrete four-component MTM, two families of symmetry 
reductions are considered. Under the local reduction $(q_2,q_4)=(q_1^*,q_3^*)$, we obtain a discrete MTM; imposing the nonlocal constraint 
$(q_2, q_4)=(-\mathrm{i}q_{1,-}^*, \mathrm{i}q_{3,-}^*)$,
yields a nonlocal discrete MTM system. Remarkably, the discrete MTM admits only AKNS-type solutions, 
whereas the nonlocal discrete MTM accommodates both AKNS-type and KP-type solutions. 
Since only nonlocal complex reductions can be realized within the Cauchy matrix reduction scheme (see \cite{XZ2022}), 
we restrict our consideration in this paper to the nonlocal complex reduction $(q_2, q_4)=(-\mathrm{i}q_{1,-}^*, \mathrm{i}q_{3,-}^*)$ 
and do not address the nonlocal real reduction $(q_2,q_4)=(q_{1,-},q_{3,-})$.

The paper is organized as follows. In Section 2, we introduce the DES and master functions $\bS^{(i,j)}$. 
Using the recurrence and shift relations for $\bS^{(i,j)}$, we construct the discrete four-component MTM in closed form. 
In Section 3, we investigate continuum limits of the discrete four-component MTM.
Section 4 is devoted to the AKNS-type and KP-type solutions 
to the resulting discrete four-component MTM by solving the DES, and Section 5 concerns its conjugate reductions. 
Conclusions and some remarks are given in Section 6.

\section{Discrete four-component massive Thirring model}

In this section, we study the discretization of the four-component MTM \eqref{eq:continuous-mtm} using the Cauchy matrix 
framework. We begin with the DES, consisting of a Sylvester matrix equation and compatible discrete shift relations, 
and recapitulate essential master functions' identities that underpin the derivation of discrete equations. Component-wise decomposition combined with 
proper variable transformations leads to the discrete four-component MTM.

Before presenting the DES, we recall the solvability condition for the Sylvester equation
\begin{align}
\label{SE}
\bX\bM-\bM\bY=\bZ,
\end{align}
where $\bX, \bY$ and $\bZ$ are known matrices and $\bM$ is the unknown matrix. For the solution of the Sylvester equation \eqref{SE}, a well known result,
which was proved by Sylvester \cite{Sylvester-1884}, is
\begin{prop}\label{prop-1}
Let us denote the eigenvalue sets of $\bX$ and $\bY$ by $\mathcal{E}(\bX)$ and $\mathcal{E}(\bY)$, respectively.
For the known $\bX, \bY$ and $\bZ$, the Sylvester equation \eqref{SE} has a unique solution $\bM$ if and only if
$\mathcal{E}(\bX)\bigcap \mathcal{E}(\bY)=\varnothing$.
\end{prop}
This result is always called the Sylvester Theorem.

\subsection{The DES and master functions}
\label{sec-2-1}

Let us start from the DES
\begin{subequations}
	\label{DES}
	\begin{align}
		\label{SE1}
		&\bK\bM-\bM\bL=\br\bs^{\st},\\
		\label{sh-r}
		& p\wt{\br}=p\br+\bK\br\ba,\quad q\wh{\br}=q\br+\bK^{-1}\br\ba,\\
		\label{sh-s}
		& p\wt{\bs}+\bL^{\st}\wt{\bs}\ba=p\bs,\quad q\wh{\bs}+(\bL^{-1})^{\st}\wh{\bs}\ba=q\bs,
	\end{align}
\end{subequations} 
where $\bK,\bL\in \mathbb{C}^{N\times N}$ are constant matrices, $\bM\in \mathbb{C}^{N\times N}[n,m]$,
and $\br, \bs \in \mathbb{C}^{N\times 2}[n,m]$ are matrix functions of $(n,m)\in \mathbb{Z}^2$;
$\ba=\mathrm{diag}(1,-1)$ stands for the third Pauli matrix $\sigma_{3}$, and the superscript $^{\st}$ denotes transposition;  $p$ and $q$ are the lattice
parameters associated with the $n$- and $m$-directions, respectively. We adopt standard shorthand notations for shifts along these two directions, namely:
$$
\wt{f}=f(n+1, m), \quad \wh{f}=f(n, m+1), \quad \wh{\wt{f}}=f(n+1, m+1).
$$

\begin{remark}
Systems \eqref{sh-r} and \eqref{sh-s} are employed to determine the discrete plane-wave factors (discrete dispersion relations), whose general form is given by
\begin{align}
\label{plane1}
\rho=\left( \frac{p+k}{p-k}\right)^{n}\left( \frac{q+k^{-1}}{q-k^{-1}}\right)^{m}\rho_0.
\end{align}
This expression differs formally from the plane-wave factor adopted in \cite{NCQ1983}, which reads
\begin{align}
\label{plane2}
\rho=\left( \theta\frac{p-k}{p^*-k}\right)^{n}\left(\theta^{'} \frac{q-k}{q^*-k}\right)^{m}\rho_0,\quad |\theta|=|\theta^{'}|=1.
\end{align}
However, by setting $p=\pm\mathrm{i}|p|=(\mathrm{i}a)^{-1}, \theta=1$ and $q=\pm\mathrm{i}|q|=\mathrm{i}b, \theta^{'}=-1$ in \eqref{plane2}, 
the plane-wave factor \eqref{plane2} reduces to the one employed in this work. This procedure is also noted in \cite{ZMZ2026}
for recovering the continuous MTM from discrete counterparts. We therefore conclude that the plane-wave factors adopted in this paper and in \cite{NCQ1983}
are essentially identical.
\end{remark}

Rather than passing immediately to scalar variables, we encode the DES in a family of master functions
\begin{equation}
	\label{Sij}
	\bS^{(i,j)}=\bs^{\st}\bL^{j}(\bC+\bM)^{-1}\bK^{i}\br, \quad i,j\in\mathbb{Z}, 
\end{equation}
where $\bC\in\mathbb{C}^{N\times N}$ is chosen such that $\bC+\bM$ is invertible. If we assume that the Sylvester equation \eqref{SE1} 
admits a unique solution $\bM$, under the commutation constraint
\begin{align}\label{kc-cl}
	\bK\bC-\bC\bL=\bm 0,
\end{align}
the master functions $\bS^{(i,j)}$ satisfy the following recurrence relations and shift relations.
\begin{itemize}
	\item
	Recurrence relations:
	\begin{subequations}
		\label{Sij-re=1}
		\begin{align}
			\label{Sij-re=1-po}
			\bS^{(i,j+1)}=&\bS^{(i+1,j)}-\bS^{(0,j)}\bS^{(i,0)},\\
			\label{Sij-re=1-ne}
			\bS^{(i,j-1)}=&\bS^{(i-1,j)}+\bS^{(-1,j)}\bS^{(i,-1)}.
		\end{align}
	\end{subequations}
	\item
	Shift relations along the $n$- and $m$-directions:
	\begin{subequations}
		\label{sh-Sij}
		\begin{align}
			\label{p-Sij-1}
			p\wt{\bS}^{(i,j)}+\ba\wt{\bS}^{(i,j+1)}
			=&p\bS^{(i,j)}+\bS^{(i+1,j)}\ba-\bS^{(0,j)}\ba\wt{\bS}^{(i,0)},\\
			\label{p-Sij-2}	
			p\bS^{(i,j)}-\ba\bS^{(i,j+1)}=&p\wt{\bS}^{(i,j)}-\wt{\bS}^{(i+1,j)}\ba+\wt{\bS}^{(0,j)}\ba\bS^{(i,0)},\\
			\label{q-Sij-1}
			q\wh{\bS}^{(i,j)}+\ba\wh{\bS}^{(i,j-1)}
			=&q\bS^{(i,j)}+\bS^{(i-1,j)}\ba+\bS^{(-1,j)}\ba\wh{\bS}^{(i,-1)},\\
			\label{q-Sij-2}
			q\bS^{(i,j)}-\ba\bS^{(i,j-1)}=&q\wh{\bS}^{(i,j)}-\wh{\bS}^{(i-1,j)}\ba-\wh{\bS}^{(-1,j)}\ba\bS^{(i,-1)}.
		\end{align}
	\end{subequations}
\end{itemize}
These relations follow directly from the DES \eqref{DES} and serve as the algebraic machinery for the subsequent construction; 
their derivations can be found in \cite{HSZ2025,ZMZ2026}.

The condition \eqref{kc-cl} admits two canonical realizations that are relevant for constructing explicit solutions. 
When $\bL=\bK$, $\bC$ commutes with $\bK$ and can be normalized to $\bI$, which yields the AKNS-type formulation. 
We denote by $\bI$ the unit matrix, with its dimension omitted.
If $\bK$ and $\bL$ share no common eigenvalues, the Sylvester theorem forces $\bC=\boldsymbol{0}$, leading to the KP-type formulation. Accordingly,
the explicit expressions for the master functions in these two schemes are given by
\begin{align}
	\bS^{(i,j)}=
	\begin{cases}
		\bs^{\st}\bK^{j}(\bI+\bM)^{-1}\bK^{i}\br , \quad \mbox{AKNS case},\\
		\bs^{\st}\bL^{j}\bM^{-1}\bK^{i}\br , \quad \mbox{KP case}.
	\end{cases}
\end{align}
In the analysis below, we always assume that $\bK$ and $\bL$ share no common eigenvalues for the KP-type scheme 
($\bK\neq\bL$), while $\bK_{1}$ and $\bK_{2}$ share no common eigenvalues in the AKNS scheme where $\bK=\bL=\mathrm{Diag}(\bK_1,\bK_2)$. 
It is convenient not to distinguish these two cases at this stage. The matrix identities we use to construct the discrete four-component 
MTM are universal and independent of the particular solution scheme adopted. We therefore postpone the distinction between the two schemes until Section 3.

\subsection{The discrete four-component MTM}

We now select three matrix-valued combinations of master functions:
\begin{align}
	\label{uvw-def}
	&\bu=\bS^{(0,0)}=\begin{pmatrix}
		u_{1}&u_{2}\\
		u_{3}&u_{4}
	\end{pmatrix}, \quad \bv=\bI-\bS^{(-1,0)}=\begin{pmatrix}
		v_{1}&v_{2}\\
		v_{3}&v_{4}
	\end{pmatrix},\quad \bw=\bI+\bS^{(0,-1)},
\end{align}
where $u_i, v_{i}~(i=1,2,3,4)$ are scalar functions.  Substituting the corresponding indices into \eqref{Sij-re=1} and \eqref{sh-Sij} gives the following matrix relations.
\begin{prop}
The matrix functions $\bu$, $\bv$ and $\bw$ introduced in \eqref{uvw-def} satisfy
	\begin{subequations}
		\label{re-uvw}
		\begin{align}
			\label{vw}
			&\bv\bw=\bI,\\
			\label{h-shift-uvw}
			&\bv\ba\wh{\bw}=\wh{\bv}\ba\bw=\ba-q(\wh{\bu}-\bu),\\
			\label{t-shift-uvw3}
			&p\bv\wt{\bw}=p\bI+\bu\ba-\ba\wt{\bu},\\
			\label{t-shift-uvw4}
			&p\wt{\bv}\bw=p\bI+\ba\bu-\wt{\bu}\ba.
		\end{align}
	\end{subequations}
\end{prop}
For the derivation of these matrix relations, one can refer to \cite{HSZ2025,ZMZ2026}.

To derive the discrete four-component MTM, in addition to the matrix-valued functions $\bu,\bv$ and $\bw$ defined earlier, 
we introduce a further quantity 
\begin{align}
\label{def-h}
\bh:=\bS^{(-1,-1)}=\begin{pmatrix}
	h_{1}&h_{2}\\h_{3}&h_{4}
\end{pmatrix},
\end{align}
and proceed to explore the relations among $\bv, \bw$ and $\bh$.
We observe that evaluating relations \eqref{p-Sij-1} and \eqref{p-Sij-2} at the specific indices $(i,j)=(-1,-1)$ yields
\begin{subequations}
\label{t-vwh}
\begin{align}
&\ba+p(\wt{\bh}-\bh)=\bw\ba\wt{\bv},\\
&\ba+p(\wt{\bh}-\bh)=\wt{\bw}\ba\bv,
\end{align}
\end{subequations}
which govern the evolution of the matrix-valued functions $\bv,\bw$ and $\bh$ in the $n$-direction. 
Specializing the index pair $(i,j)$ to $(-1,0)$ in equations \eqref{q-Sij-1} and \eqref{q-Sij-2} and noticing $\bS^{(-2,0)}=\bv\bh$,
we perform elementary algebraic rearrangement and arrive at the $m$-directional 
evolution relations for $\bv,\bw$ and $\bh$:
\begin{subequations}
\label{h-vwh}
\begin{align}
&q\bw\wh{\bv}=q\bI+\ba\wh{\bh}-\bh\ba,\\
&q\wh{\bw}\bv=q\bI+\wh{\bh}\ba-\ba\bh.
\end{align}
\end{subequations}

Thus far, we have established several key shift relations: the $m$-directional evolution relations for $\bu,\bv$ and $\bw$ given in equation
\eqref{h-shift-uvw}; the $n$-directional evolution relations for $\bu,\bv$ and $\bw$ presented in equations \eqref{t-shift-uvw3} 
and \eqref{t-shift-uvw4}; the $n$-directional evolution relations for $\bv,\bw$ and $\bh$ in equation \eqref{t-vwh}; 
and the $m$-directional evolution relations for $\bv,\bw$ and $\bh$ in equation \eqref{h-vwh}. These relations constitute 
the fundamental basis for the derivation of the discrete four-component MTM. We now proceed to further manipulate these evolution relations to
construct this model.

We perform a component-wise expansion of the matrix-valued evolution relations to recast them in scalar form. 
As a prerequisite, we first determine the four entries of matrix $\bw$. Since $\bw$ is the inverse of $\bv$, one can normalize $\bv$ and $\bw$ by
\begin{align}
	\label{trans1}
	\bV= \frac{1}{\sqrt{|\bv|}}\bv=\begin{pmatrix}
		V_{1}&V_{2}\\
		V_{3}&V_{4}
	\end{pmatrix}, \quad \bW= \frac{1}{\sqrt{|\bw|}}\bw
\end{align}
and consider $\bV$ and $\bW$, or instead, still employ the $\bv$ and $\bw$ but assume $|\bv| =1$. We prefer the later, i.e.  
assuming  $|\bv| =1$ without loss of generality (justified in \cite{LQZ2023,ZMZ2026}). Thus, the four entries of $\bw$ can be explicitly expressed as
\begin{align}
	\label{w-def1}
	\bw=\bI+\bS^{(0,-1)}=\begin{pmatrix}
		v_{4}&-v_{2}\\
		-v_{3}&v_{1}
	\end{pmatrix}.
\end{align}
We proceed to decouple the matrix evolution relations \eqref{h-shift-uvw}, \eqref{t-shift-uvw3}, \eqref{t-shift-uvw4}, \eqref{t-vwh} and \eqref{h-vwh} 
into relations satisfied by the scalar-valued functions $u_{i},v_{i}$ and $h_{i}$ for $i=1,2,3,4$. The complete set of resulting scalar relations is summarized as follows.
\begin{itemize}
\item
Component-wise expansion of matrix equation \eqref{h-shift-uvw}:
\begin{subequations}
\label{com-wi-1}
\begin{align}
&1-q(\wh{u}_{1}-u_{1})=v_{1}\wh{v}_{4}+v_{2}\wh{v}_{3},\\
&q(\wh{u}_{2}-u_{2})=v_{1}\wh{v}_{2}+\wh{v}_{1}v_{2},\\
\label{hu3}
&q(\wh{u}_{3}-u_{3})=-v_{3}\wh{v}_{4}-\wh{v}_{3}v_{4},\\
\label{hu4}
&v_{1}\wh{v}_{4}+v_{2}\wh{v}_{3}=\wh{v}_{1}v_{4}+\wh{v}_{2}v_{3};
\end{align}
\end{subequations}
\item
Component-wise expansion of matrix equations \eqref{t-shift-uvw3} and \eqref{t-shift-uvw4}:
\begin{subequations}
\label{com-wi-2}
\begin{align}
&p+u_{1}-\wt{u}_{1}=p(v_{1}\wt{v}_{4}-v_{2}\wt{v}_{3}),\\
&u_{2}+\wt{u}_{2}=p(v_{1}\wt{v}_{2}-\wt{v}_{1}v_{2}),\\
\label{tu3}
&u_{3}+\wt{u}_{3}=p(v_{3}\wt{v}_{4}-\wt{v}_{3}v_{4}),\\
\label{tu4}
&\wt{v}_{1}v_{4}-\wt{v}_{2}v_{3}=v_{1}\wt{v}_{4}-v_{2}\wt{v}_{3};
\end{align}
\end{subequations}
\item
Component-wise expansion of matrix equation \eqref{t-vwh}:
\begin{subequations}
\label{com-wi-3}
\begin{align}
&1+p(\wt{h}_{1}-h_{1})=\wt{v}_{1}v_{4}+v_{2}\wt{v}_{3},\\
\label{th2}
&p(\wt{h}_{2}-h_{2})=v_{2}\wt{v}_{4}+\wt{v}_{2}v_{4},\\
&p(\wt{h}_{3}-h_{3})=-v_{1}\wt{v}_{3}-\wt{v}_{1}v_{3};
\end{align}
\end{subequations}
\item
Component-wise expansion of matrix equation \eqref{h-vwh}:
\begin{subequations}
\label{com-wi-4}
\begin{align}
&q-h_{1}+\wh{h}_{1}=q(\wh{v}_{1}v_{4}-v_{2}\wh{v}_{3}),\\
\label{hh2}
&h_{2}+\wh{h}_{2}=q(\wh{v}_{2}v_{4}-v_{2}\wh{v}_{4}),\\
&h_{3}+\wh{h}_{3}=q(\wh{v}_{1}v_{3}-v_{1}\wh{v}_{3}),
\end{align}
\end{subequations}
\end{itemize}
where in \eqref{com-wi-3} and \eqref{com-wi-4}, we have omitted one relation each, since they are nothing but \eqref{hu4} and \eqref{tu4}, respectively.

The scalar functions $u_{i},v_{i}$ and $h_{i}$ admit diverse algebraic combinations, some of which form closed discrete equations that 
are discrete counterparts of celebrated integrable equations. For instance, $u_{2}$ and $u_{3}$ give the discrete negative-order AKNS 
equation \cite{Z2016}, while $u_{3}$ and $\mathrm{i}v_{2}/v_{4}$ (or $\mathrm{i}v_{3}/v_{1}$ and $u_{2}$) yield the discrete Mikhailov model \cite{ZMZ2026}. 

We next introduce four new variables:
\begin{align}
\label{q1q2q3q4}
q_{1}:=v_{3},\quad q_{2}:=-\mathrm{i}h_{2}/v_{4},\quad q_{3}:=\mathrm{i}u_{3}/v_{4},\quad q_{4}:=v_{2}.
\end{align}
With $q_{1},q_{2},q_{3}$ and $q_{4}$ defined as above, the discrete four-component MTM is obtained with these quantities 
as the dependent variables. The detailed derivation is presented as follows.

By virtue of the determinant normalization condition for matrix \(\bv\), namely \(v_{1}v_{4}-v_{2}v_{3}=1\), we have
\begin{align}
v_{1}=(q_{1}q_{4}+1)/v_{4}.
\end{align}
Substituting this expression into equations \eqref{hu4} and \eqref{tu4} and carrying out algebraic rearrangements, we arrive at
\begin{subequations}
\label{q1q4ab}
\begin{align}
&(\wh{q}_{1}+\alpha q_{1})(\wh{q}_{4}-\alpha q_{4})=\alpha^2-1,\\
&(\wt{q}_{1}-\beta q_{1})(\wt{q}_{4}+\beta q_{4})=\beta^2-1,
\end{align}
\end{subequations}
which, with the help of relations \eqref{q1q2q3q4ab-2} and \eqref{q1q2q3q4ab-3}, are further transformed into
\begin{subequations}
\label{q1q2a-q3q4b}
\begin{align}
&\mathrm{i}(\wh{q}_{1}+\alpha q_{1})(q_{2}+\alpha \wh{q}_{2})=q(\alpha^2-1),\\
&\mathrm{i}(q_{3}+\beta \wt{q}_{3})(\wt{q}_{4}+\beta q_{4})=p(\beta^2-1),
\end{align}
\end{subequations}
where $\alpha=\wh{v}_{4}/v_{4}$ and $\beta=\wt{v}_{4}/v_{4}$.
Equation \eqref{q1q2a-q3q4b} involves the intermediate parameters $\alpha$ and $\beta$, 
which need to be eliminated by substituting identities expressed in terms of $q_{1}, q_{2}, q_{3}$ and $q_{4}$.
Rewriting relations \eqref{hu3}, \eqref{hh2}, \eqref{tu3} and \eqref{th2} under the substitutions
$v_{3}=q_{1}$, $h_{2}=\mathrm{i}v_{4}q_{2}$, $u_{3}=-\mathrm{i}v_{4}q_{3}$ and $v_{2}=q_{4}$, we obtain
\begin{subequations}
\label{q1q2q3q4ab}
\begin{align}
\label{q1q2q3q4ab-1}
&\mathrm{i}q(q_{3}-\alpha\wh{q}_{3})+\wh{q}_{1}+\alpha q_{1}=0,\\
\label{q1q2q3q4ab-2}
&\mathrm{i}q(\wh{q}_{4}-\alpha q_{4})+q_{2}+\alpha\wh{q}_{2}=0,\\
\label{q1q2q3q4ab-3}
&\mathrm{i}p(\wt{q}_{1}-\beta q_{1})+q_{3}+\beta\wt{q}_{3}=0,\\
\label{q1q2q3q4ab-4}
&\mathrm{i}p(q_{2}-\beta\wt{q}_{2})+\wt{q}_{4}+\beta q_{4}=0.
\end{align}
\end{subequations}
These relations provide identities that express $\alpha$ as a function of $q_{1},q_{2}$ and $\beta$ as a function of $q_{3},q_{4}$.
Using the identities in \eqref{q1q2q3q4ab}, we eliminate the intermediate parameters $\alpha$ and $\beta$ from \eqref{q1q2a-q3q4b}, 
and finally obtain the lattice system for the variables $q_{1}, q_{2}, q_{3}$ and $q_{4}$:
\begin{subequations}
\label{unreduced-DMTM}
\begin{align}
&(q_{1}-\mathrm{i}q\wh{q}_{3})^2-(\wh{q}_{1}+\mathrm{i}qq_{3})^2+(q_{1}q_{3}+\wh{q}_{1}\wh{q}_{3})[q_{2}(q_{1}-\mathrm{i}q\wh{q}_{3})-\wh{q}_{2}(\wh{q}_{1}+\mathrm{i}qq_{3})]=0,\\
&(q_{2}+\mathrm{i}q\wh{q}_{4})^2-(\wh{q}_{2}-\mathrm{i}qq_{4})^2+(q_{2}q_{4}+\wh{q}_{2}\wh{q}_{4})[q_{1}(q_{2}+\mathrm{i}q\wh{q}_{4})-\wh{q}_{1}(\wh{q}_{2}-\mathrm{i}qq_{4})]=0,\\
&(q_{3}+\mathrm{i}p\wt{q}_{1})^2-(\wt{q}_{3}-\mathrm{i}pq_{1})^2+(q_{1}q_{3}+\wt{q}_{1}\wt{q}_{3})[q_{4}(q_{3}+\mathrm{i}p\wt{q}_{1})-\wt{q}_{4}(\wt{q}_{3}-\mathrm{i}pq_{1})]=0,\\
&(q_{4}-\mathrm{i}p\wt{q}_{2})^2-(\wt{q}_{4}+\mathrm{i}pq_{2})^2+(q_{2}q_{4}+\wt{q}_{2}\wt{q}_{4})[q_{3}(q_{4}-\mathrm{i}p\wt{q}_{2})-\wt{q}_{3}(\wt{q}_{4}+\mathrm{i}pq_{2})]=0,
\end{align}
\end{subequations}
We refer to this system as the discrete four-component MTM, since it reduces to the continuous four-component MTM \eqref{eq:continuous-mtm} 
under the appropriate continuum limit (see Section \ref{sec-cl}).

\begin{remark}
As an alternative to the substitution prescribed in \eqref{q1q2q3q4}, the ansatz
\begin{align}
	\label{q1q2q3q4-1}
	 q_{1}:=-\mathrm{i}h_{3}/v_{1},\quad q_{2}:=v_{2},\quad q_{3}:=v_{3}, \quad q_{4}:=\mathrm{i}u_{2}/v_{1},
\end{align}
also satisfies the discrete four-component MTM \eqref{unreduced-DMTM}.
\end{remark}

In \cite{WLWZ2026}, the relation between the dependent variables of the Mikhailov model and the four-component MTM model is 
presented explicitly (see Proposition 1 therein). The result below establishes the connection between the dependent variables of 
the discrete Mikhailov model given in \cite{ZMZ2026} 
and those of the discrete four-component MTM model constructed in the present paper.
\begin{theorem}
Let $(\mu,\nu)$ denote the solutions of the discrete Mikhailov model (see Eq. (2.57) in \cite{ZMZ2026}), 
and $(q_1, q_2, q_3, q_4)$ be the solutions of the discrete four-component MTM \eqref{unreduced-DMTM}. 
Within the Cauchy matrix framework, these quantities are defined respectively as
\begin{align}
\label{uvq1q2q3q4}
\mu:=u_{3},\quad\nu:=\mathrm{i}v_{2}/v_{4},\quad q_{1}:=v_{3},\quad q_{2}:=-\mathrm{i}h_{2}/v_{4},\quad q_{3}:=\mathrm{i}u_{3}/v_{4},\quad q_{4}:=v_{2},
\end{align}
where $u_{i}$, $v_{i}$ and $h_{i}$ are specified in \eqref{uvw-def} and \eqref{def-h}. It then follows that
\begin{subequations}
\begin{align}
\label{q1-uv}
&p(\wt{q}_{1}v_{4}-q_{1}\wt{v}_{4})+\mu+\wt{\mu}=0,\quad \wh{q}_{1}v_{4}+q_{1}\wh{v}_{4}+q(\wh{\mu}-\mu)=0,\\
\label{q2-uv}
&p(\wt{q}_{2}\wt{v}_{4}-q_{2}v_{4})+v_{4}\wt{v}_{4}(\nu+\wt{\nu})=0,\quad \wh{q}_{2}\wh{v}_{4}+q_{2}v_{4}+qv_{4}\wh{v}_{4}(\wh{\nu}-\nu)=0,\\
\label{q3q4-uv}
&q_{3}v_{4}-\mathrm{i}\mu=0,\quad q_{4}+\mathrm{i}\nu v_{4}=0,
\end{align}
\end{subequations}
and $v_{4}$ satifies
\begin{align}
\label{re-v4-munu}
\wh{v}_4/v_4-v_{4}/\wh{v}_{4}
	=\mathrm{i} q(\wh\mu-\mu)(\wh\nu-\nu),\quad p(\wt{v}_4/v_4-v_{4}/\wt{v}_{4}) 
	=\mathrm{i}(\mu+\wt\mu)(\nu+\wt\nu).
\end{align}
\end{theorem}
\begin{proof}
Relation \eqref{q3q4-uv} follows immediately from the definitions of $\mu$, $\nu$, $q_3$ and $q_4$ in \eqref{uvq1q2q3q4}. 
We now turn to the verification of relations \eqref{q1-uv} and \eqref{q2-uv}. Substituting $u_{3}=\mu$ and $v_{3}=q_{1}$ into 
\eqref{hu3} and \eqref{tu3} directly yields \eqref{q1-uv}, whereas \eqref{q2-uv} is obtained by inserting $h_{2}=\mathrm{i}q_{2}v_{4}$ and $v_{2}=-\mathrm{i}\nu v_{4}$
into \eqref{th2} and \eqref{hh2}.

Set $\alpha=\wh{v}_4/v_4$ and $\beta=\wt{v}_{4}/v_4$, as introduced in \eqref{q1q2q3q4ab}. The constraint $|\bv|=1$ gives
\begin{align}
v_1=(1+v_2v_3)/v_4,\quad \wh{v}_1=(1+\wh{v}_2\wh{v}_3)/\wh{v}_4,\quad
\wt{v}_1=(1+\wt{v}_2\wt{v}_3)/\wt{v}_4.
\end{align}
Substituting the identities derived above into \eqref{hu4} and \eqref{tu4} and performing algebraic rearrangement leads to
\begin{subequations}
\label{v3v2}
\begin{align}
&(\wh{v}_3+\alpha v_3)(\wh{v}_2-\alpha v_2)=\alpha^2-1,\\
&(\wt{v}_3-\beta v_3)(\wt{v}_2+\beta v_2)=\beta^2-1.
\end{align}
\end{subequations}
The quantities $\wh{v}_2-\alpha v_2$ and $\wt{v}_2+\beta v_2$ obey the identities
\begin{align}
\label{v2v4v}
\wh{v}_2-\alpha v_2=-\mathrm{i}\alpha v_4(\wh{\nu}-\nu),\quad \wt{v}_2+\beta v_2=-\mathrm{i}\beta v_4(\wt{\nu}+\nu),
\end{align}
Combining these with \eqref{hu3} and \eqref{tu3}, we further obtain
\begin{align}
\label{v3v4u}
\wh{v}_3+\alpha v_3=-q(\wh{\mu}-\mu)/v_4, \quad \wt{v}_3-\beta v_3=-(\mu+\wt{\mu})/(pv_4).
\end{align}
Inserting \eqref{v2v4v} and \eqref{v3v4u} into \eqref{v3v2} produces
\begin{subequations}
\begin{align}
&\alpha-\alpha^{-1}
=\mathrm{i} q(\widehat\mu-\mu)(\widehat\nu-\nu),\\
&p(\beta-\beta^{-1})
=\mathrm{i}(\mu+\widetilde\mu)(\nu+\widetilde\nu),
\end{align}
\end{subequations}
which is exactly the relation \eqref{re-v4-munu}.
This completes the proof.
\end{proof}

\section{Continuum limits of the discrete four-component MTM \eqref{eq:continuous-mtm}}
\label{sec-cl}

In this section, we verify that the discrete four-component MTM \eqref{unreduced-DMTM} recovers the four-component continuous MTM via two-step continuum limits. 

In the first step, we take $q\rightarrow\infty$, $m\rightarrow\infty$
with the continuous spatial coordinate $x=m/q$ kept finite. The $m$-direction shift is expanded as
\begin{align}
\label{f-h}
\wh{f}=f(n,x+1/q)=f+f_{x}/q+f_{xx}/(2q^2)+\cdots.
\end{align}
We reformulate the expressions of $\wh{q}_{1},\wh{q}_{2},\wh{q}_{3}$ and $\wh{q}_{4}$ in accordance with the general expanded form 
presented in \eqref{f-h}, and substitute the resulting expressions into the discrete four-component MTM \eqref{unreduced-DMTM}. 
By expanding in powers of $1/q$, we extract the leading-order coefficient of order $\mathcal{O}(q)$, which yields the semi-discrete four-component MTM:
\begin{subequations}
\label{sdMTM}
\begin{align}
&\mathrm{i}q_{3,x}-2q_{1}-2q_{1}q_{2}q_{3}=0,\\
&\mathrm{i}q_{4,x}+2q_{2}+2q_{1}q_{2}q_{4}=0,\\
&(q_{3}+\mathrm{i}p\wt{q}_{1})^2-(\wt{q}_{3}-\mathrm{i}pq_{1})^2+(q_{1}q_{3}+\wt{q}_{1}\wt{q}_{3})[q_{4}(q_{3}+\mathrm{i}p\wt{q}_{1})-\wt{q}_{4}(\wt{q}_{3}-\mathrm{i}pq_{1})]=0,\\
&(q_{4}-\mathrm{i}p\wt{q}_{2})^2-(\wt{q}_{4}+\mathrm{i}pq_{2})^2+(q_{2}q_{4}+\wt{q}_{2}\wt{q}_{4})[q_{3}(q_{4}-\mathrm{i}p\wt{q}_{2})-\wt{q}_{3}(\wt{q}_{4}+\mathrm{i}pq_{2})]=0.
\end{align}
\end{subequations}
It is worth noting that the dependent variables $q_{1}, q_{2}, q_{3}$ and $q_{4}$ now depend on both the discrete 
coordinate $n$ and the continuous coordinate $x$, namely $q_{1}:=q_{1}(n,x),~~q_{2}:=q_{2}(n,x),~~q_{3}:=q_{3}(n,x),~~q_{4}:=q_{4}(n,x)$. 

\begin{theorem}\label{Theor-sd}
Let $(\mu,\nu)$ denote the solutions of the semi-discrete Mikhailov model (see Eq. (4.4) in \cite{ZMZ2026}), 
and $(q_1, q_2, q_3, q_4)$ be the solutions of the semi-discrete four-component MTM \eqref{sdMTM}. 
All of these quantities depend on the discrete variable $n$ and the continuous variable $x$, and their definitions are 
formally identical to those in \eqref{uvq1q2q3q4}. It follows that
\begin{align}
\label{re-q1234-munu-semi}
q_{1}=-\mu_{x}/(2v_{4}),\quad q_{2}=-v_{4}\nu_{x}/2,\quad q_{3}=\mathrm{i}\mu/v_{4},\quad q_{4}=-\mathrm{i}v_{4}\nu,
\end{align}
where $v_{4}$ satifies
\begin{align}
v_{4,x}/v_{4}=\mathrm{i}\mu_{x}\nu_{x}/2 \quad \Longrightarrow \quad v_{4}=\exp(\frac{\mathrm{i}}{2}\partial^{-1}
(\mu_{x}\nu_{x})).
\end{align}
Here $\partial^{-1}=\frac{1}{2}\left(\int_{-\infty}^x-\int_{x}^{+\infty}\right)\cdot \dd x$ (cf. \cite{ZMZ2026}).
\end{theorem}
Relation \eqref{re-q1234-munu-semi} shares a similar structure with Equation (2.13) in \cite{WLWZ2026}.

In the second step, we take $p\rightarrow\infty$ and $n\rightarrow\infty$, while keeping the continuous temporal coordinate $t=n/p$ finite.
The $n$-direction shift is expanded as
\begin{align}
	\wt{f}=f(t+1/p,x)=f+f_{t}/p+f_{tt}/(2p^2)+\cdots.
\end{align}
Substituting the expanded expressions of $\wt{q}_{1},\wt{q}_{2},\wt{q}_{3}$ and $\wt{q}_{4}$ into the semi-discrete four-component MTM \eqref{sdMTM} 
and retaining the leading-order terms (in terms of $\mathcal{O}(p)$), we recover the system
\begin{subequations}
\label{MTM}
\begin{align}
&\mathrm{i}q_{3,x}-2q_{1}-2q_{1}q_{2}q_{3}=0,\\
&\mathrm{i}q_{4,x}+2q_{2}+2q_{1}q_{2}q_{4}=0,\\
&\mathrm{i}q_{1,t}+2q_{3}+2q_{1}q_{3}q_{4}=0,\\
&\mathrm{i}q_{2,t}-2q_{4}-2q_{2}q_{3}q_{4}=0,
\end{align}
\end{subequations}
which is nothing but the celebrated four-component MTM \eqref{eq:continuous-mtm} after a coordinate scaling $(t,x)\rightarrow(t/2,-x/2)$.

\begin{remark}
If we regard the dependent variables $\mu,\nu,q_1, q_2, q_3, q_4$ in Theorem \ref{Theor-sd} as functions of the continuous 
independent variables $t$ and $x$, the conclusion of Theorem \ref{Theor-sd} yields the relation between solutions of the 
continuous Mikhailov equation (see Eq. (4.10) in \cite{ZMZ2026}) and the continuous four-component MTM \eqref{MTM}.
\end{remark}

\section{Some examples of exact solutions}
\label{sec-solu}

The DES \eqref{DES} has been solved in \cite{ZMZ2026}. To facilitate the discussion in Section \ref{sec-redu}, 
we nevertheless present the main results for the solutions of the discrete four-component MTM \eqref{unreduced-DMTM}.

We first define the discrete plane-wave factors:
\begin{subequations}
	\label{plane-def}
	\begin{align}
		\label{plane-rho-def}
		&\rho_{\iota}(z)=\left( 1+(-1)^{\iota-1}\frac{z}{p}\right)^{n}
		\left(1+ (-1)^{\iota-1}\frac{1}{qz}\right)^{m}\rho^{0}_{\iota}(z),\\
		&\sigma_{\iota}(z)=\left( 1+(-1)^{\iota-1}\frac{z}{p}\right)^{-n}
		\left(1+ (-1)^{\iota-1}\frac{1}{qz}\right)^{-m}\sigma^{0}_{\iota}(z),
	\end{align}
\end{subequations}
where $\rho^{0}_{\iota}(z),~\sigma^{0}_{\iota}(z),~(\iota=1,2)$ are constant phase parameters independent of $n$ and $m$.

For the sake of simplicity, we list the Weinstein--Aronszajn formula 
\begin{align}
\label{GDI}
|\bI+\mbx\mby|=|\bI+\mby\mbx|,
\end{align}
where $\mbx$ and $\mby$ are $N\times M$ and $M\times N$ matrices. 

\subsection{AKNS-type solutions}
For the AKNS-type scheme ($\bK=\bL, N=2 \mathcal{N}$ ), we introduce block-structured matrices
\begin{align}
	\label{Sym-KM12rts-def}
	\bK=\begin{pmatrix}
		\bK_{1}&\\
		&\bK_{2}
	\end{pmatrix}, \quad \bM=\begin{pmatrix}
		&\bM_{1}\\
		\bM_{2}&
	\end{pmatrix}, \quad \br=\begin{pmatrix}
		\br_{1}&\\
		&\br_{2}
	\end{pmatrix}, \quad \bs=\begin{pmatrix}
		& \bs_{1} \\
		\bs_{2} &
	\end{pmatrix},
\end{align}	
where $\bK_{\iota}, \bM_{\iota} \in \mathbb{C}^{\mathcal{N}\times \mathcal{N}}$, and $\br_{\iota},\bs_{\iota}$ are $\mathcal{N}$-dimensional column vectors 
for $\iota=1,2$. Under the assumption that $\bK_1$ and $\bK_2$ have no common eigenvalues, the DES \eqref{DES} decouples into
\begin{subequations}
	\label{DES-sym-ent}
	\begin{align}
		\label{sym-SE-12}
		& \bK_{1}\bM_{1}-\bM_{1}\bK_{2}=\br_{1}\bs^{\st}_{2},\quad \bK_{2}\bM_{2}-\bM_{2}\bK_{1}=\br_{2}\bs^{\st}_{1}, \\
		\label{sym-shift-r1}
		& p\wt{\br}_{\iota}=p\br_{\iota}+(-1)^{\iota-1}\bK_{\iota}\br_{\iota},\quad q\wh{\br}_{\iota}=q\br_{\iota}+(-1)^{\iota-1}\bK^{-1}_{\iota}\br_{\iota}, \\
		\label{sym-shift-s1}
		& p\wt{\bs}_{\iota}+(-1)^{\iota}\bK^{\st}_{\iota}~\wt{\bs}_{\iota}=p\bs_{\iota},\quad q\wh{\bs}_{\iota}
+(-1)^{\iota}(\bK_{\iota}^{-1})^{\st}~\wh{\bs}_{\iota}=q\bs_{\iota}.
	\end{align}
\end{subequations}
The matrix function $\bv$ defined in equation \eqref{uvw-def} is formulated as
\begin{align}
\label{sym-v-def}
\bv=\bI-\bS^{(-1,0)}=\bI-\bs^{\st}(\bI+\bM)^{-1}\bK^{-1}\br.
\end{align}
We proceed to evaluate the determinant of $\bv$. Performing elementary algebraic simplification gives rise to
\begin{align}
	\label{sym-det-v}
	|\bv| &=\left|\bI-\bs^{\st}(\bI+\bM)^{-1}\bK^{-1}\br\right| \nn \\
	&= \left|\bI-(\bI+\bM)^{-1}\bK^{-1}\br\bs^{\st}\right| \nn \\
	&=\left|(\bI+\bM)^{-1}\bK^{-1}(\bI+\bM)\bK\right| \nn \\
	&=1,
\end{align}
where we have used the Weinstein--Aronszajn formula.
This is consistent with the normalization assumption $|\bv|=1$ introduced in Section 2.2.
Thus, the AKNS-type solutions $q_{1}, q_{2}, q_{3}$ and $q_{4}$ of the discrete four-component MTM \eqref{unreduced-DMTM} are expressed as
\begin{subequations}
\label{AKNS-solu}
\begin{align}
\label{AKNS-q1}
&q_{1}=-\bs_{1}^{\st}(\bI-\bM_{1}\bM_{2})^{-1}\bK_{1}^{-1}\br_{1},\\
\label{AKNS-q2}
&q_{2}=\frac{-\mathrm{i}\bs_{2}^{\st}\bK_{2}^{-1}(\bI-\bM_{2}\bM_{1})^{-1}\bK_{2}^{-1}\br_{2}}{1+\bs_{1}^{\st}\bM_{1}(\bI-\bM_{2}\bM_{1})^{-1}\bK_{2}^{-1}\br_{2}},\\
\label{AKNS-q3}
&q_{3}=\frac{\mathrm{i}\bs_{1}^{\st}(\bI-\bM_{1}\bM_{2})^{-1}\br_{1}}{1+\bs_{1}^{\st}\bM_{1}(\bI-\bM_{2}\bM_{1})^{-1}\bK_{2}^{-1}\br_{2}},\\
\label{AKNS-q4}
&q_{4}=-\bs_{2}^{\st}(\bI-\bM_{2}\bM_{1})^{-1}\bK_{2}^{-1}\br_{2},
\end{align}
\end{subequations}
where the involved elements are determined by the system \eqref{DES-sym-ent}.

In what follows, we derive the explicit soliton solutions and multiple-pole solutions of the discrete four-component
MTM \eqref{unreduced-DMTM} by solving the system \eqref{DES-sym-ent}.

\vskip 5pt
\noindent
\textbf{Soliton solutions}. To get the soliton solutions, we take
\begin{align}
	\label{K12-def}
	\bK_{1}=\text{diag}(k_{1},k_{2},\cdots,k_{\mathcal{N}}), \quad \bK_{2}=\text{diag}(l_{1},l_{2},\cdots,l_{\mathcal{N}}),
\end{align}
with $k_i\neq l_j,~(i,j=1,2,\ldots,\mathcal{N}$). Then we have
\begin{subequations}
	\label{sym-SE-solu}
	\begin{align}
		&\br_{1}=(\rho_{1}(k_{1}),\rho_{1}(k_{2}),\cdots,\rho_{1}(k_{\mathcal{N}}))^{\st}, \quad 
\br_{2}=(\rho_{2}(l_{1}),\rho_{2}(l_{2}),\cdots,\rho_{2}(l_{\mathcal{N}}))^{\st},\\
		&\bs_{1}=(\sigma_{2}(k_{1}),\sigma_{2}(k_{2}),\cdots,\sigma_{2}(k_{\mathcal{N}}))^{\st}, \quad 
\bs_{2}=(\sigma_{1}(l_{1}),\sigma_{1}(l_{2}),\cdots,\sigma_{1}(l_{\mathcal{N}}))^{\st},\\
		\label{AKNS-M1-def}
		&\bM_{1}=(M_{1,ij})_{\mathcal{N}\times \mathcal{N}},\quad M_{1,ij}=\frac{\rho_{1}(k_{i})\sigma_{1}(l_{j})}{k_{i}-l_{j}},\\
		\label{AKNS-M2-def}
		&\bM_{2}=(M_{2,ij})_{\mathcal{N}\times \mathcal{N}},\quad M_{2,ij}=\frac{\rho_{2}(l_{i})\sigma_{2}(k_{j})}{l_{i}-k_{j}}.
	\end{align}
\end{subequations}
Substituting \eqref{K12-def} and \eqref{sym-SE-solu} into \eqref{AKNS-solu} leads to $\mathcal{N}$-th soliton solutions to the discrete four-component MTM \eqref{unreduced-DMTM}.

\vskip 5pt
\noindent
\textbf{Multiple-pole solutions}. Solutions of this type
arise from the Jordan-block structure of matrices $\bK_1$ and $\bK_2$.
Here, we consider two forms of the matrices
$\bK_{1}$ and $\bK_{2}$.

\noindent$\bullet$\hspace{1em}\textit{Case-I}. We take $\bK_{1}$ as a Jordan-block matrix and
$\bK_{2}$ as a Jordan-block-like matrix
\begin{align}
	\label{K12-Jor}
	\bK_{1}=\bJ[k,1], \quad \bK_{2}=\bJ[l,-1], \quad \text{with} \quad k\neq l,
\end{align}
where
\begin{align}
	\label{J-def}
	&\bJ[z,\delta]:=\begin{pmatrix}
		z&0&0&\cdots&0\\
		\delta&z&0&\cdots&0\\
		0&\delta&z&\cdots&0\\
		\vdots&\vdots&\vdots&\ddots&\vdots\\
		0&0&0&\cdots&z
	\end{pmatrix}_{\mathcal{N}\times \mathcal{N}}.
\end{align}
In this situation, we have
\begin{subequations}
	\label{JS-sym}
	\begin{align}
		&\br_{1}=\bF_{k}[\rho_{1}(k)]\bar{\be}_{\mathcal{N}},\quad \br_{2}=\bF_{-l}[\rho_{2}(l)]\bar{\be}_{\mathcal{N}}, \\
		&\bs_{1}=\bH_{k}[\sigma_{2}(k)]\bar{\be}_{\mathcal{N}},\quad
		\bs_{2}=\bH_{-l}[\sigma_{1}(l)]\bar{\be}_{\mathcal{N}},\\
		&\bM_{1}=\bF_{k}[\rho_{1}(k)]\bar{\bG}\bH_{-l}[\sigma_{1}(l)],\quad
		\bM_{2}=-\bF_{-l}[\rho_{2}(l)]\bar{\bG}\bH_{k}[\sigma_{2}(k)], \\
		&\bar{\bG}=(\bar{G}_{i,j})_{\mathcal{N}\times \mathcal{N}},\quad \bar{G}_{i,j}=\frac{(-1)^{i+j}
			\mathrm{C}^{i-1}_{i+j-2}}{(k-l)^{i+j-1}},\quad \bar{\be}_{\mathcal{N}}=(\underbrace{1, 0, \cdots, 0}_{\mathcal{N}\text{-dimensional}})^{\st},
	\end{align}
\end{subequations}
where $\mathrm{C}^{i}_{j}=\frac{j!}{i!(j-i)!}$ for $j\geq i$ and
\begin{subequations}
	\label{FH-def}
	\begin{align}
		\label{F-def}
		&\bF_{\delta z}[f(z)]=\begin{pmatrix}
			f(z)&0&0&\cdots&0\\
			\partial_{\delta z}f(z)&f(z)&0&\cdots&0\\
			\frac{\partial^{2}_{\delta z}f(z)}{2!}&\partial_{\delta z}f(z)&f(z)
			&\cdots&0\\
			\vdots&\vdots&\vdots&\ddots&\vdots\\
			\frac{\partial^{\mathcal{N}-1}_{\delta z}f(z)}{(\mathcal{N}-1)!}
			&\frac{\partial^{\mathcal{N}-2}_{\delta z}f(z)}{(\mathcal{N}-2)!}
			&\frac{\partial^{\mathcal{N}-3}_{\delta z}f(z)}{(\mathcal{N}-3)!}&\cdots&f(z)
		\end{pmatrix}_{\mathcal{N}\times \mathcal{N}},\\
		\label{H-def}
		&\bH_{\delta z}[f(z)]=
		\begin{pmatrix}
			\frac{\partial^{\mathcal{N}-1}_{\delta z}f(z)}{(\mathcal{N}-1)!} & \cdots 
			& \frac{\partial^{2}_{\delta z}f(z)}{2!} & \partial_{\delta z}f(z) & f(z)\\
			\frac{\partial^{\mathcal{N}-2}_{\delta z}f(z)}{(\mathcal{N}-2)!} 
			& \cdots & \partial_{\delta z}f(z) & f(z) & 0\\
			\frac{\partial^{\mathcal{N}-3}_{\delta z}f(z)}{(\mathcal{N}-3)!} & \cdots & f(z) & 0 & 0\\
			\vdots & \vdots & \vdots & \ddots &\ddots\\
			f(z) & \cdots & 0 & 0 & 0
		\end{pmatrix}_{\mathcal{\mathcal{N}}\times \mathcal{\mathcal{N}}}.
	\end{align}
\end{subequations}
Substituting \eqref{K12-Jor} and \eqref{JS-sym} into \eqref{AKNS-solu} leads to
$\mathcal{N}$-th order multiple-pole solutions to the discrete four-component MTM \eqref{unreduced-DMTM}.

\noindent$\bullet$\hspace{1em}\textit{Case-II}. We take
\begin{align}
	\bK_{1}=\bJ[k,1],\quad \bK^{\st}_{2}=\bJ[l,1].
\end{align}
In this case, we have
\begin{subequations}
	\begin{align}
		&\br_{1}=\bF_{k}[\rho_{1}(k)]\bar{\be}_{\mathcal{N}},\quad \br_{2}=\bH_{l}[\rho_{2}(l)]\bar{\be}_{\mathcal{N}},\\
		&\bs_{1}=\bH_{k}[\sigma_{2}(k)]\bar{\be}_{\mathcal{N}},\quad
		\bs_{2}=\bF_{l}[\sigma_{1}(l)]\bar{\be}_{\mathcal{N}},\\
		&\bM_{1}=\bF_{k}[\rho_{1}(k)]\bG_{1}\bF^{\st}_{l}[\sigma_{1}(l)],
		\quad\bM_{2}=\bH_{l}[\rho_{2}(l)]\bG_{2}\bH_{k}[\sigma_{2}(k)],\\
		&\bG_{1}=(g^{(1)}_{i,j})_{\mathcal{N}\times \mathcal{N}},\quad g^{(1)}_{i,j}=\frac{(-1)^{i-1}\mathrm{C}^{i-1}_{i+j-2}}{(k-l)^{i+j-1}},\\
		&\bG_{2}=(g^{(2)}_{i,j})_{\mathcal{N}\times \mathcal{N}},\quad g^{(2)}_{i,j}=\frac{(-1)^{i-1}\mathrm{C}^{i-1}_{i+j-2}}{(l-k)^{i+j-1}},
	\end{align}
\end{subequations}
which yield another $\mathcal{N}$-th order multiple-pole solutions to \eqref{unreduced-DMTM}.

\begin{remark}
At this point, we clarify the rationale for presenting two distinct forms of the multiple-pole solution above. 
This distinction stems from their different behaviors under reductions in Section \ref{sec-redu}. Specifically, the multiple-pole solution of the 
first form reduces to the multiple-pole solution of the discrete MTM \eqref{DMTM}, whereas the second form yields the multiple-pole 
solution of the nonlocal discrete MTM \eqref{non-DMTM} upon reduction.
\end{remark}

\subsection{KP-type solutions}

For the KP-type scheme~($\bK\neq \bL,\bC=\bm 0$), we decompose $\br=(\br_{1},\br_{2})$, $\bs=(\bs_{1},\bs_{2})$ and $\bM=\bM_{1}+\bM_{2}$, 
leading to two decoupled scalar KP-type Cauchy matrix systems:
\begin{subequations}
	\label{asym-shift1}
	\begin{align}
		\label{asym-SE12}
		& \bK\bM_{\iota}-\bM_{\iota}\bL=\br_{\iota}\bs^{\st}_{\iota}, \\
		\label{asym-shift-r1}
		& p\wt{\br}_{\iota}=p\br_{\iota}+(-1)^{\iota-1}\bK\br_{\iota},\quad q\wh{\br}_{\iota}=q\br_{\iota}+(-1)^{\iota-1}\bK^{-1}\br_{\iota}, \\
		\label{asym-shift-s1}
		& p\wt{\bs}_{\iota}+(-1)^{\iota-1}\bL^{\st}~\wt{\bs}_{\iota}=p\bs_{\iota},\quad
		q\wh{\bs}_{\iota}+(-1)^{\iota-1}(\bL^{-1})^{\st}~\wh{\bs}_{\iota}=q\bs_{\iota}.
	\end{align}
\end{subequations}
The matrix function $\bv$ defined in \eqref{uvw-def} reads
\begin{align}
	\label{asmy-v-def}
	\bv=\bI-\bS^{(-1,0)}=\bI-\bs^{\st}\bM^{-1}\bK^{-1}\br.
\end{align}
Since
\begin{align}
	\label{asym-det-v}
	\left|\bv\right|=\left|\bI-\bs^{\st}\bM^{-1}\bK^{-1}\br \right|
	=\left|\bL\right|/\left|\bK\right|,
\end{align}
which is a constant (maybe not 1), then, by virtue of the normalization transformation \eqref{trans1}, 
the KP-type solutions $q_{1}, q_{2}, q_{3}$ and $q_{4}$ of the four-component DMTM \eqref{unreduced-DMTM} are given by
\begin{subequations}
\label{KP-solu}
\begin{align}
\label{KP-q1}
&q_{1}=V_{3}=\tau v_{3}=-\tau\bs^{\st}_{2}(\bM_{1}+\bM_{2})^{-1}\bK^{-1}\br_{1},\\
\label{KP-q2}
&q_{2}=-\mathrm{i}\frac{h_{2}}{V_{4}}=-\mathrm{i}\frac{h_{2}}{\tau v_{4}}=\frac{-\mathrm{i}\bs_{1}^{\st}\bL^{-1}
(\bM_{1}+\bM_{2})^{-1}\bK^{-1}\br_{2}}{\tau(1-\bs^{\st}_{2}(\bM_{1}+\bM_{2})^{-1}\bK^{-1}\br_{2})},\\
\label{KP-q3}
&q_{3}=\mathrm{i}\frac{u_{3}}{V_{4}}=\mathrm{i}\frac{u_{3}}{\tau v_{4}}=\frac{\mathrm{i}\bs^{\st}_{2}
(\bM_{1}+\bM_{2})^{-1}\br_{1}}{\tau(1-\bs^{\st}_{2}(\bM_{1}+\bM_{2})^{-1}\bK^{-1}\br_{2})},\\
\label{KP-q4}
&q_{4}=V_{2}=\tau v_{2}=-\tau\bs^{\st}_{1}(\bM_{1}+\bM_{2})^{-1}\bK^{-1}\br_{2},
\end{align}
where the parameter $\tau$ satisfies $\tau^2=|\bK|/|\bL|$.
\end{subequations}

Solving system \eqref{asym-shift1} yields the explicit expression for \eqref{KP-solu}, which describes both the multi-soliton and multiple-pole 
solutions of the discrete four-component MTM \eqref{unreduced-DMTM}. The explicit forms of the constituent elements in these solutions are presented below.
\vskip 5pt
\noindent
\textbf{Soliton solutions}. 
\begin{subequations}
	\label{asym-SE-solu}
	\begin{align}
		& \bK=\text{diag}(k_{1},k_{2},\cdots,k_{N}),\quad \bL=\text{diag}(l_{1},l_{2},\cdots,l_{N}),\\
		\label{asym-M1M2-decom}
		&\br_{\iota}=(\rho_{\iota}(k_{1}),\rho_{\iota}(k_{2}),\cdots,\rho_{\iota}(k_{N}))^{\st},\quad 
\bs_{\iota}=(\sigma_{\iota}(l_{1}),\sigma_{\iota}(l_{2}),\cdots,\sigma_{\iota}(l_{N}))^{\st},\\
		&\bM_{\iota}=(M_{\iota,ij})_{N\times N},\quad M_{\iota,ij}=\frac{\rho_{\iota}(k_{i})\sigma_{\iota}(l_{j})}{k_{i}-l_{j}},
	\end{align}
\end{subequations}
with $k_i\neq l_j,~(i,j=1,2,\ldots,N$), in which $\rho_{\iota},\sigma_{\iota}~(\iota=1,2)$ are defined in \eqref{plane-def}.

\vskip 5pt
\noindent
\textbf{Multiple-pole solutions}.
\begin{subequations}
	\begin{align}
		&\bK=\bJ[k,1],\quad \bL^{\st}=\bJ[l,1],\\
		&\br_{\iota}=\bF_{k}[\rho_{\iota}(k)]\bar{\be}_{N},\quad
		\bs_{\iota}=\bF_{l}[\sigma_{\iota}(l)]\bar{\be}_{N}, \quad \bar{\be}_{N}
		=(\underbrace{1, 0, \cdots, 0}_{N\text{-dimensional}})^{\st}, \\
		&\bM=\bM_{1}+\bM_{2},\quad\bM_{\iota}
		=\bF_{k}[\rho_{\iota}(k)]\bar{\bG}\bF^{\st}_{l}[\sigma_{\iota}(l)],
		\quad \iota=1,2,\\
		&\bar{\bG}=(\bar{G}_{i,j})_{N\times N},\quad \bar{G}_{i,j}=\frac{(-1)^{i-1}\mathrm{C}^{i-1}_{i+j-2}}{(k-l)^{i+j-1}},
	\end{align}
where the matrices $\bJ$ and $\bF$ are defined by \eqref{J-def} and \eqref{F-def}, respectively.
\end{subequations}

\section{Local and nonlocal reductions}
\label{sec-redu}

In this section, we impose two distinct reduction constraints to derive, respectively, the standard local discrete MTM and the reverse-$(n,m)$ nonlocal discrete MTM.
To construct solutions of the local and nonlocal discrete MTM, appropriate reduction constraints must be imposed on the spectral and Cauchy matrix data.
Instead of solving the reduced discrete equations, we perform reductions directly at the level of the algebraic solution ansatz:
we derive conditions on the matrix quantities $\bK, \bL, \br, \bs, \bM$ such that the unreduced solution automatically satisfies either 
the local complex-conjugation reduction or the nonlocal conjugation reduction.
In what follows, we first address the local reduction within the AKNS-type framework, before proceeding to the nonlocal reduction for both 
AKNS-type and KP-type realizations. All these results can be adapted to the semi-discrete setting via the continuum limits introduced in Section \ref{sec-cl}.

\subsection{Local reduction}

By imposing reduction $(q_{2}, q_{4})=(q_{1}^*, q_{3}^*)$, one immediately gets a local discrete MTM
\begin{subequations}
\label{DMTM}
\begin{align}
&(q_{1}-\mathrm{i}q\wh{q}_{3})^2-(\wh{q}_{1}+\mathrm{i}qq_{3})^2+(q_{1}q_{3}+\wh{q}_{1}
\wh{q}_{3})[q_{1}^*(q_{1}-\mathrm{i}q\wh{q}_{3})-\wh{q}_{1}^*(\wh{q}_{1}+\mathrm{i}qq_{3})]=0,\\
&(q_{3}+\mathrm{i}p\wt{q}_{1})^2-(\wt{q}_{3}-\mathrm{i}pq_{1})^2+(q_{1}q_{3}+\wt{q}_{1}
\wt{q}_{3})[q_{3}^*(q_{3}+\mathrm{i}p\wt{q}_{1})-\wt{q}_{3}^*(\wt{q}_{3}-\mathrm{i}pq_{1})]=0.
\end{align}
\end{subequations}

With the solutions of the discrete four-component MTM derived in Section \ref{sec-solu}, we proceed to identify matrix constraints 
under which the obtained solutions obey the local complex conjugation reduction automatically. 
Only the AKNS-type scheme is admitted in this section.

\begin{theorem}\label{Theo-AKNS-redu}
	The AKNS-type solution \eqref{AKNS-solu} admits local reduction $(q_{2}, q_{4})=(q_{1}^*, q_{3}^*)$ under constraints
	\begin{align}
		\label{AKNS-redu1}
		\bK_{2}=-\bK_{1}^{*},\quad \br_{2}=\mathrm{i}\bK_{2}\br_{1}^{*},\quad \bs_{2}=\bs_{1}^{*}.
	\end{align}
\end{theorem}
\begin{proof}
By condition \eqref{AKNS-redu1}, from \eqref{DES-sym-ent} one immediately obtains
\begin{align}
\bK_{1}\bM_{1}+\bM_{1}\bK_{1}^{*}=\br_{1}\bs_{1}^{\dagger},\quad
\bK_{1}(\mathrm{i}\bK_{1}^{-1}\bM_{2}^{*})+(\mathrm{i}\bK_{1}^{-1}\bM_{2}^{*})\bK_{1}^{*}=\br_{1}\bs_{1}^{\dagger},
\end{align}
where notation $^{\dagger}$ represents the Hermitian adjoint.
Subtracting the second equation from the first one and using the Sylvester Theorem, 
one easily finds the relations between $\bM_{1}$ and $\bM_{2}$: 
\begin{align}
	\label{M1M2-re}
	\bM_{2}^{*}=-\mathrm{i}\bK_{1}\bM_{1},\quad\bM_{1}^{*}=\mathrm{i}\bK_{2}^{-1}\bM_{2}\quad\Longrightarrow\quad\bM_{1}^{*}\bM_{2}^{*}=\bK_{2}^{-1}\bM_{2}\bK_{1}\bM_{1}.
\end{align}
To simplify subsequent calculations, define
\begin{align}
\label{ABt}
\bA:=\bI-\bM_{1}\bM_{2}, \quad \bB:=\bI-\bM_{2}\bM_{1}, \quad \bt:=\bK_{2}^{-1}\br_{2},
\end{align}
and denote the common denominator of $q_{2}, q_{3}$ by
\begin{align}
\label{v4-def}
v_{4}=1+\bs_{1}^{\st}\bM_{1}\bB^{-1}\bt.
\end{align}
Then the AKNS-type solution \eqref{AKNS-solu} can be written as
\begin{align}
	\label{re-def-q1q2q3q4}
	q_{1}=-\bs_{1}^{\st}\bA^{-1}\bK_{1}^{-1}\br_{1},\quad q_{2}=-\mathrm{i}\dfrac{\bs_{2}^{\st}\bK_{2}^{-1}\bB^{-1}\bt}{v_{4}},\quad
	q_{3}=\mathrm{i}\dfrac{\bs_{1}^{\st}\bA^{-1}\br_{1}}{v_{4}},\quad q_{4}=-\bs_{2}^{\st}\bB^{-1}\bt.
\end{align}
We next carry out the proof in three steps.

\textbf{Step 1:} We verify that the common denominator $v_{4}$ satisfies the identity $v_{4}v_{4}^*=1$. 
It follows directly from the second equation of \eqref{sym-SE-12} that
\begin{align}
\label{step1-1}
\bK_{2}^{-1}\bM_{2}\bK_{1}=\bM_{2}-\bt\bs_{1}^{\st}.
\end{align}
Combining this relation with equation \eqref{M1M2-re}, we obtain
\begin{align}
\label{step1-2}
\bA^{*}=\bI-\bM_{1}^{*}\bM_{2}^{*} =\bI-\bK_{2}^{-1}\bM_{2}\bK_{1}\bM_{1} =\bB+\bt\bs_{1}^{\st}\bM_{1}.
\end{align}
Using the Weinstein--Aronszajn formula, we find
\begin{align}
|\bA^{*}|=|\bB| \cdot \left(1+\bs_{1}^{\st}\bM_{1}\bB^{-1}\bt\right)=|\bB| \cdot v_{4}.
\end{align}
Furthermore, since
\begin{align}
|\bB|=|\bI-\bM_{2}\bM_{1}|=|\bI-\bM_{1}\bM_{2}|=|\bA|,
\end{align}
it follows that
\begin{align}
\label{step1-5}
v_{4}=|\bA^{*}|/|\bA| \Longrightarrow v_{4}v_{4}^{*}=1.
\end{align}

\textbf{Step 2}: Proof of $q_{2}=q_{1}^*$.
From equations \eqref{M1M2-re} and \eqref{step1-1},
\begin{align}
\label{step2-1}
\bK_{1}^{-1}\bB^{*}\bK_{1}=\bI-\bM_{1}\bK_{2}^{-1}\bM_{2}\bK_{1}
=\bA+\bM_{1}\bt\bs_{1}^{\st}.
\end{align}
Let $\bX$ denote the last matrix in equation \eqref{step2-1}. From $\bA\bM_{1}=\bM_{1}\bB$, we have
\begin{align}
\bA^{-1}\bM_{1}=\bM_{1}\bB^{-1}.
\end{align}
Therefore,
\begin{align}
\bs_{1}^{\st}\bA^{-1}\bX =\bs_{1}^{\st}+(\bs_{1}^{\st}\bA^{-1}\bM_{1}\bt)\bs_{1}^{\st}
=(1+\bs_{1}^{\st}\bM_{1}\bB^{-1}\bt)\bs_{1}^{\st}=v_{4}\bs_{1}^{\st}.
\end{align}
Right-multiplying both sides by $\bX^{-1}$ and dividing by $v_{4}$, we obtain
\begin{align}
\label{step2-4}
\bs_{1}^{\st}\bX^{-1}=\bs_{1}^{\st}\bA^{-1}/v_{4}.
\end{align}
On the other hand, condition \eqref{AKNS-redu1} gives
\begin{align}
\left(\bK_{2}^{-1}\right)^{*}=-\bK_{1}^{-1}, \quad\bt^{*}=-\mathrm{i}\br_{1}.
\end{align}
Taking the complex conjugate of $q_{2}$ in equation \eqref{re-def-q1q2q3q4}, and using $\bK_{1}^{-1}(\bB^{*})^{-1}=\bX^{-1}\bK_{1}^{-1}$, 
as well as the equations \eqref{step1-5} and \eqref{step2-4}, we have
\begin{align}
q_{2}^{*} =-\frac{\bs_{1}^{\st}\bK_{1}^{-1}\left(\bB^{*}\right)^{-1}\br_{1}}{v_{4}^{*}}
=-\frac{\bs_{1}^{\st}\bX^{-1}\bK_{1}^{-1}\br_{1}}{v_{4}^{*}} 
=-\frac{\bs_{1}^{\st}\bA^{-1}\bK_{1}^{-1}\br_{1}}{v_{4} v_{4}^{*}}
=-\bs_{1}^{\st}\bA^{-1}\bK_{1}^{-1}\br_{1}=q_{1}.
\end{align}

\textbf{Step 3}: Proof of $q_{3}=q_{4}^{*}$. From \eqref{step1-2}, one has
\begin{align}
\bA^*\bB^{-1}\bt=(\bB+\bt\bs_{1}^{\st}\bM_{1})\bB^{-1}\bt=\bt(1+\bs_{1}^{\st}\bM_{1}\bB^{-1}\bt)=v_{4}\bt.
\end{align}
Thus
\begin{align}
(\bA^{*})^{-1}\bt=\bB^{-1}\bt/v_{4}.
\end{align}
Using $\br_{1}^{*}=-\mathrm{i}\bt$, $\bs_{1}^{*}=\bs_{2}$ and equation \eqref{step1-5}, a direct calculation yields
\begin{align}
q_{3}^{*}=-\frac{\mathrm{i} \bs_{2}^{\st}\left(\bA^{*}\right)^{-1} \br_{1}^{*}}{v_{4}^{*}} 
=-\frac{\bs_{2}^{\st}\left(\bA^{*}\right)^{-1} \bt}{v_{4}^{*}} 
=-\frac{\bs_{2}^{\st} \bB^{-1} \bt}{v_{4} v_{4}^{*}} 
=-\bs_{2}^{\st} \bB^{-1} \bt=q_{4}.
\end{align}
The proof is completed.
\end{proof}

The following theorem reveals the AKNS-type Cauchy matrix soliton solutions and multiple-pole solutions to the discrete MTM \eqref{DMTM}.
\begin{theorem}\label{Theo-AKNS-solu}
The functions
\begin{subequations}
\begin{align}
&q_{1}=-\bs_{1}^{\st}(\bI-\mathrm{i}\bM_{1}\bK_{1}^*\bM_{1}^*)^{-1}\bK_{1}^{-1}\br_{1},\\
&q_{3}=\frac{\mathrm{i}\bs_{1}^{\st}(\bI-\mathrm{i}\bM_{1}\bK_{1}^*\bM_{1}^*)^{-1}\br_{1}}
{1+\mathrm{i}\bs_{1}^{\st}\bM_{1}(\bI-\mathrm{i}\bK_{1}^*\bM_{1}^*\bM_{1})^{-1}\br_{1}^*},
\end{align}
\end{subequations}
provide i) the AKNS-type Cauchy matrix $\mathcal{N}$-soliton solutions for the discrete MTM \eqref{DMTM} with
	\begin{subequations}
		\label{AKNS-msr-def}
		\begin{align}
			\label{AKNS-k-def}
			&\bK_{1}=\mathrm{diag}(k_{1},k_{2},\cdots,k_{\mathcal{N}}),\quad k_{i}\in\mathbb{C}, \quad i=1,2,\cdots,\mathcal{N},\\
			\label{AKNS-r-def}
			&\bs_{1}=(\sigma_{2}(k_{1}),\sigma_{2}(k_{2}),\cdots,\sigma_{2}(k_{\mathcal{N}}))^{\st},\quad 
			\br_{1}=(\rho_{1}(k_{1}),\rho_{1}(k_{2}),\cdots,\rho_{1}(k_{\mathcal{N}}))^{\st},\\
			\label{AKNS-m1-def}
			&\bM_{1}=(M_{1,ij})_{\mathcal{N}\times \mathcal{N}},\quad M_{1,ij}=\frac{\rho_{1}(k_{i})\sigma_{2}^{*}(k_{j})}{k_{i}+k_{j}^{*}};
		\end{align}
	\end{subequations}
and ii) the AKNS-type Cauchy matrix $\mathcal{N}$-pole solutions for the discrete MTM \eqref{DMTM} with
	\begin{subequations}
		\label{AKNS-msr-def-1}
		\begin{align}
			&\bK_{1}=\bJ[k,1],\quad\br_{1}=\bF_{k}[\rho_{1}(k)]\bar{\be}_{N},\quad\bs_{1}=\bH_{k}[\sigma_{2}(k)]\bar{\be}_{N},\\
			&\bM_{1}=\bF_{k}[\rho_{1}(k)]\bar{\bG}\bH^{*}_{k}[\sigma_{2}(k)], \quad
			\bar{\bG}=(\bar{G}_{i,j})_{\mathcal{N}\times \mathcal{N}},\quad \bar{G}_{i,j}=\frac{(-1)^{i+j}\mathrm{C}^{i-1}_{i+j-2}}{(k+k^{*})^{i+j-1}},
		\end{align}
	\end{subequations}
where the discrete plane-wave factors $\rho_{1}(z)$ and $\sigma_{2}(z)$ are defined in \eqref{plane-def},
and the matrices $\bJ,\bF,\bH$ are defined in \eqref{J-def} and \eqref{FH-def}, respectively.
\end{theorem}

\begin{remark}

Under the constraint $(q_{2}, q_4)=(q_{1}^*,q_{3}^*)$, the semi-discrete four-component MTM \eqref{sdMTM} and the four-component MTM \eqref{MTM}
reduce to the semi-discrete MTM
\begin{subequations}
\label{sdMTM-local}
\begin{align}
&\mathrm{i}q_{3,x}-2q_{1}-2|q_{1}|^2q_{3}=0,\\
&(q_{3}+\mathrm{i}p\wt{q}_{1})^2-(\wt{q}_{3}-\mathrm{i}pq_{1})^2+(q_{1}q_{3}+\wt{q}_{1}\wt{q}_{3})
[q_{3}^*(q_{3}+\mathrm{i}p\wt{q}_{1})-\wt{q}_{3}^*(\wt{q}_{3}-\mathrm{i}pq_{1})]=0,
\end{align}
\end{subequations}
and the MTM
\begin{subequations}
\label{MTM-local}
\begin{align}
&\mathrm{i}q_{3,x}-2q_{1}-2|q_{1}|^2q_{3}=0,\\
&\mathrm{i}q_{1,t}+2q_{3}+2q_{1}|q_{3}|^2=0,
\end{align}
\end{subequations}
respectively. Theorem \ref{Theo-AKNS-solu} also provides the AKNS-type Cauchy matrix $\mathcal{N}$-soliton solutions and 
AKNS-type Cauchy matrix $\mathcal{N}$-pole solutions for these reduced systems, 
upon replacing $\rho_{\iota}$ and  $\sigma_{\iota}$ with semi-discrete plane-wave factors: 
\begin{subequations}
	\label{semi-plane-def}
	\begin{align}
		&\varrho_{\iota}(z)=\left( 1+(-1)^{\iota-1}\frac{z}{p}\right)^{n}\exp((-1)^{\iota-1}x/z)\rho^{0}_{\iota}(z),\\
		&\chi_{\iota}(z)=\left(1+(-1)^{\iota-1}\frac{z}{p}\right)^{-n}
		\exp((-1)^{\iota}x/z)\sigma^{0}_{\iota}(z),
	\end{align}
\end{subequations}
 as well as with continuous exponential functions:
\begin{subequations}
\label{con-plane-def}
\begin{align}
& \theta_{\iota}(z)=\exp((-1)^{\iota-1}(zt+x/z))\rho^{0}_{\iota}(z),\\
& \vartheta_{\iota}(z)=\exp((-1)^{\iota}(zt+x/z))\sigma^{0}_{\iota}(z).
\end{align}
\end{subequations}
\end{remark}

\subsection{Nonlocal reduction}

We now address the nonlocal reduction associated with the reversed lattice. 
Upon imposing the nonlocal reduction $(q_{2}, q_{4})=(-\mathrm{i}q^*_{1,-}, \mathrm{i}q^*_{3,-})$,
the discrete four-component MTM \eqref{unreduced-DMTM} is reduced to a nonlocal discrete MTM:
\begin{subequations}
	\label{non-DMTM}
	\begin{align}
		&(q_{1}-\mathrm{i}q\wh{q}_{3})^2-(\wh{q}_{1}+\mathrm{i}qq_{3})^2+\mathrm{i}(q_{1}q_{3}+\wh{q}_{1}\wh{q}_{3})
[\wh{q}^*_{1,-}(\wh{q}_{1}+\mathrm{i}qq_{3})-q^*_{1,-}(q_{1}-\mathrm{i}q\wh{q}_{3})]=0,\\
		&(q_{3}+\mathrm{i}p\wt{q}_{1})^2-(\wt{q}_{3}-\mathrm{i}pq_{1})^2+\mathrm{i}(q_{1}q_{3}+\wt{q}_{1}\wt{q}_{3})
[q_{3,-}^*(q_{3}+\mathrm{i}p\wt{q}_{1})-\wt{q}_{3,-}^*(\wt{q}_{3}-\mathrm{i}pq_{1})]=0.
	\end{align}
\end{subequations}

Unlike the local case,  both the AKNS-type and KP-type Cauchy matrix schemes satisfy this nonlocal constraint, and we discuss them in turn.

\begin{theorem}\label{Th-AK-non}
The AKNS-type solution \eqref{AKNS-solu} admits reduction relation $(q_{2}, q_{4})=(-\mathrm{i}q^*_{1,-}, \mathrm{i}q^*_{3,-})$, provided that
the elements satisfy the constraints
\begin{align}
		\label{AKNS-nonlocal-redu1}
		\bK_{2}=\bK_{1}^{\dagger}, \quad \br_{2}=\bK_{2}(\bs^{\st}_{1,-})^{\dagger}, \quad \bs^{\st}_{2}=-\br_{1,-}^{\dagger}.
\end{align}
\end{theorem}
\begin{proof}
By \eqref{AKNS-nonlocal-redu1}, we observe that the Sylvester equation \eqref{sym-SE-12} becomes
\begin{align}
	\label{nl-sym-Sy}		
	\bK_{1}\bM_{1}-\bM_{1}\bK_{1}^{\dagger}=-\br_{1}\br_{1,-}^{\dagger},\quad\bK_{1}^{\dagger}\bM_{2}-\bM_{2}\bK_{1}
	=\bK_{1}^{\dagger}(\bs^{\st}_{1,-})^{\dagger}\bs^{\st}_{1}.
\end{align} 
Taking the Hermitian conjugate of the two above equations yields
\begin{align}
	\label{nl-sym-Sy-1}	
	\bM^{\dagger}_{1,-}\bK_{1}^{\dagger}-\bK_{1}\bM^{\dagger}_{1,-}=-\br_{1}\br_{1,-}^{\dagger},\quad
	\bM^{\dagger}_{2,-}\bK_{1}-\bK_{1}^{\dagger}\bM^{\dagger}_{2,-}=(\bs^{\st}_{1,-})^{\dagger}\bs^{\st}_{1}\bK_{1}.
\end{align}
Comparing \eqref{nl-sym-Sy} and \eqref{nl-sym-Sy-1}	and noticing the Sylvester Theorem, we derive
\begin{align}
\bM_{2,-}^{\dagger}=-(\bK_{1}^{\dagger})^{-1}\bM_{2}\bK_{1},\quad\bM_{1,-}^{\dagger}
	=-\bM_{1}\quad\Longrightarrow\quad\bM_{2,-}^{\dagger}\bM_{1,-}^{\dagger}=\bK_{2}^{-1}\bM_{2}\bK_{1}\bM_{1}.
\end{align}	
Using definitions of $\bA,\bB,\bt$ and $v_4$ in~\eqref{ABt} and \eqref{v4-def}, then by
the second Sylvester equation in \eqref{sym-SE-12} we know
\begin{align}
\bK_2^{-1}\bM_2\bK_1=\bM_2-\bt\bs_1^{\st},
\end{align}
and consequently have
\begin{align}
\bA_{-}^{\dagger}=\bI-\bM_{2,-}^{\dagger}\bM_{1,-}^{\dagger}=\bI-\bK_2^{-1}\bM_2\bK_1\bM_1=\bB+\bt\bs_1^{\st}\bM_1.
\end{align}
Applying the Weinstein--Aronszajn formula leads to
\begin{align}
|\bA_{-}^{\dagger}|
=|\bB|\bigl(1+\bs_1^{\st}\bM_1\bB^{-}\bt\bigr)
=|\bB|\cdot v_4.
\end{align}
Noticing $|\bB|=|\bA|$ and $|\bA_{-}^{\dagger}|=|\bA_{-}^*|$, we obtain
\begin{align}
v_4=|\bA_{-}^*|/|\bA|,\quad v_{4,-}^*=|\bA|/|\bA_{-}^*|,\quad\Longrightarrow\quad v_4v_{4,-}^*=1.
\end{align}
Furthermore,
\begin{align}
\bA_{-}^{\dagger}\bB^{-1}\bt
	=(\bB+\bt\bs_1^{\st}\bM_1)\bB^{-1}\bt
	=\bt\bigl(1+\bs_1^{\st}\bM_1\bB^{-1}\bt\bigr)
	=v_4\bt,
\end{align}
so that
\begin{align}
(\bA_{-}^{\dagger})^{-1}\bt=\bB^{-1}\bt/v_4.
\end{align}
Since $q_1$ is scalar, its complex conjugate coincides with its Hermitian adjoint. Thus, by \eqref{AKNS-q1},
\begin{align}
q_{1,-}^*
=-\br_{1,-}^{\dagger}\bK_2^{-1}
(\bA_{-}^{\dagger})^{-1}(\bs_{1,-}^{\st})^{\dagger}
=\bs_2^{\st}\bK_2^{-1}(\bA_{-}^{\dagger})^{-1}\bt
=\bs_2^{\st}\bK_2^{-1}\bB^{-1}\bt/v_4.
\end{align}
Comparison with \eqref{AKNS-q2} gives
\begin{align}
q_2=-\mathrm{i}q_{1,-}^*.
\end{align}
Similarly, using \eqref{AKNS-q3}, we find
\begin{align}
q_{3,-}^*
=-\frac{\mathrm{i}\,\br_{1,-}^{\dagger}
(\bA_{-}^{\dagger})^{-1}(\bs_{1,-}^{\st})^{\dagger}}
{v_{4,-}^*}
=\frac{\mathrm{i}\bs_2^{\st}(\bA_{-}^{\dagger})^{-1}\bt}
{v_{4,-}^*}
=\frac{\mathrm{i}\bs_2^{\st}\bB^{-1}\bt}{v_4v_{4,-}^*}
=\mathrm{i}\bs_2^{\st}\bB^{-1}\bt.
\end{align}
It follows from \eqref{AKNS-q4} that
\begin{align}
\mathrm{i}q_{3,-}^*=-\bs_2^{\st}\bB^{-1}\bt=q_4.
\end{align}
Therefore the verification is finished. 
\end{proof}

\begin{theorem}
\label{Theo-AKNS-solu-non}
The functions
\begin{subequations}
\begin{align}
&q_{1}=-\bs_{1}^{\st}(\bI-\bM_{1}\bM_{2})^{-1}\bK_{1}^{-1}\br_{1},\\
&q_{3}=\frac{\mathrm{i}\bs_{1}^{\st}(\bI-\bM_{1}\bM_{2})^{-1}\br_{1}}{1+\bs_{1}^{\st}\bM_{1}(\bI-\bM_{2}\bM_{1})^{-1}(\bs_{1,-}^{\st})^{\dagger}},
\end{align}
\end{subequations}
provide i) the AKNS-type Cauchy matrix $\mathcal{N}$-soliton solutions for the nonlocal discrete MTM \eqref{non-DMTM} with
\begin{subequations}
\label{AKNS-nonlocal-msr-def}
\begin{align}
			\label{AKNS-non-k-def}
			&\bK_{1}=\mathrm{diag}(k_{1},k_{2},\cdots,k_{\mathcal{N}}),\quad k_{i}\in\mathbb{C},\quad i=1,2,\cdots,\mathcal{N},\\
			\label{AKNS-non-r-def}
			&\bs_{1}=(\sigma_{2}(k_{1}),\sigma_{2}(k_{2}),\cdots,\sigma_{2}(k_{\mathcal{N}}))^{\st},\quad\br_{1}=(\rho_{1}(k_{1}),\rho_{1}(k_{2}),
			\cdots,\rho_{1}(k_{\mathcal{N}}))^{\st},\\
			\label{AKNS-non-m1-def}
			&\bM_{1}=(M_{1,ij})_{\mathcal{N}\times \mathcal{N}},\quad M_{1,ij}=\frac{-\rho_{1}(k_{i})\rho_{1,-}^{*}(k_{j})}{k_{i}-k_{j}^{*}},\\
			\label{AKNS-non-m2-def}
			&\bM_{2}=(M_{2,ij})_{\mathcal{N}\times \mathcal{N}},\quad M_{2,ij}=\frac{k_{i}^{*}\sigma_{2,-}^{*}(k_{i})\sigma_{2}(k_{j})}{k_{i}^{*}-k_{j}},
\end{align}
\end{subequations}
and ii) the AKNS-type Cauchy matrix $\mathcal{N}$-pole solution for the nonlocal discrete MTM \eqref{non-DMTM} with
\begin{subequations}
\begin{align}
			&\bK_{1}=\bJ[k,1],\quad\br_{1}=\bF_{k}[\rho_{1}(k)]\bar{\be}_{\mathcal{N}},\quad\bs_{1}=\bH_{k}[\sigma_{2}(k)]\bar{\be}_{\mathcal{N}},\\
			&\bM_{1}=-\bF_{k}[\rho_{1}(k)]\bG_1\bF^{\dagger}_{k}[\rho_{1,-}(k)],\quad\bM_{2}=\bK_{1}^{\dagger}
			\bH^{*}_{k}[\sigma_{2,-}(k)]\bG_2\bH^{\st}_{k}[\sigma_{2}(k)],\\
			&\bG_{1}=(g^{(1)}_{i,j})_{\mathcal{N}\times \mathcal{N}},\quad g^{(1)}_{i,j}=\frac{(-1)^{i-1}\mathrm{C}^{i-1}_{i+j-2}}{(k-k^*)^{i+j-1}},\\
			&\bG_{2}=(g^{(2)}_{i,j})_{\mathcal{N}\times \mathcal{N}},\quad g^{(2)}_{i,j}=\frac{(-1)^{i-1}\mathrm{C}^{i-1}_{i+j-2}}{(k^*-k)^{i+j-1}},
\end{align}
\end{subequations}
where the discrete plane-wave factors $\rho_{1}(z)$ and $\sigma_{2}(z)$ are defined as in equation \eqref{plane-def},  
and the matrix functions $\bJ,\bF,\bH$ are defined in \eqref{J-def} and \eqref{FH-def}, respectively.
\end{theorem}

\begin{theorem}
	The KP-type solution \eqref{KP-solu} admits nonlocal reduction $(q_{2}, q_{4})=(\mathrm{i}q^*_{3,-}, -\mathrm{i}q^*_{1,-})$ under the constraints	 
	\begin{align}
		\label{KP-nonlocal-redu1}
		\bL=\bK^{\dagger},\quad\br_{2}=\bK(\bs^{\st}_{2,-})^{\dagger},\quad\bs^{\st}_{1}=\br_{1,-}^{\dagger}.
	\end{align}
\end{theorem}
\begin{proof}
By analogy with the proof employed in Theorem \ref{Th-AK-non}, we derive
\begin{align}	
\bM_{1,-}^{\dagger}=-\bM_{1},\quad\bM_{2,-}^{\dagger}=-\bK^{-1}\bM_{2}\bK^{\dagger}.
\end{align}
Since $\bM=\bM_1+\bM_2$ and
$\bK^{-1}\bM_2\bL=\bM_2-\bK^{-1}\br_2\bs_2^{\st}$, it follows that
\begin{align}
\bM_{-}^{\dagger}=-\bM+\bK^{-1}\br_2\bs_2^{\st}.
\end{align}
The unnormalized entry $v_4$ appearing in \eqref{KP-solu} is
\begin{align}
v_4=1-\bs_2^{\st}\bM^{-1}\bK^{-1}\br_2.
\end{align}
For the $N\times N$ matrix $\bM$, the Weinstein--Aronszajn formula yields
\begin{align}
	|\bM_{-}^*|&=|\bM_{-}^{\dagger}| \nn \\
	&=(-1)^{N}|\bM-\bK^{-1}\br_2\bs_2^{\st}| \nn \\
	&=(-1)^{N}|\bM|\bigl(1-\bs_2^{\st}\bM^{-1}\bK^{-1}\br_2\bigr) \nn \\
	&=(-1)^{N}|\bM|\cdot v_4.
\end{align}
Thus,
\begin{align}
v_4=(-1)^{N}|\bM_{-}^*|/|\bM|, \quad v_{4,-}^*=(-1)^{N}|\bM|/|\bM_{-}^*|,\quad\Longrightarrow\quad v_4v_{4,-}^*=1.
\end{align}
In addition, the normalization in \eqref{KP-solu} satisfies
\begin{align}
\tau^2=|\bK|/|\bL|=|\bK|/|\bK^*|.
\end{align}
Consequently, $|\tau|=1$ and
\begin{align}
\tau^*=\tau^{-1}.
\end{align}
This identity holds for either fixed choice of the constant
square root $\tau$.

Multiplying the identity for $\bM_{-}^{\dagger}$ by
$\bM^{-1}\bK^{-1}\br_2$ gives
\begin{align}
\bM_{-}^{\dagger}\bM^{-1}\bK^{-1}\br_2
	&=(-\bM+\bK^{-1}\br_2\bs_2^{\st})\bM^{-1}\bK^{-1}\br_2 \nn \\
	&=-\bK^{-1}\br_2
	\bigl(1-\bs_2^{\st}\bM^{-1}\bK^{-1}\br_2\bigr) \nn \\
	&=-v_4\bK^{-1}\br_2.
\end{align}
It follows that
\begin{align}
(\bM_{-}^{\dagger})^{-1}\bK^{-1}\br_2
=-\bM^{-1}\bK^{-1}\br_2/v_4.
\end{align}
Taking the complex conjugate of \eqref{KP-q1} at $(-n,-m)$,
and using the preceding identities, we find
\begin{align}
	\begin{aligned}
	q_{1,-}^*
	&=-\tau^*\br_{1,-}^{\dagger}\bL^{-1}
	(\bM_{-}^{\dagger})^{-1}(\bs_{2,-}^{\st})^{\dagger}
	=-\tau^*\bs_1^{\st}\bL^{-1}
	(\bM_{-}^{\dagger})^{-1}\bK^{-1}\br_2\\
	&=\frac{\tau^*\bs_1^{\st}\bL^{-1}\bM^{-1}\bK^{-1}\br_2}{v_4}
	=\frac{\bs_1^{\st}\bL^{-1}\bM^{-1}\bK^{-1}\br_2}{\tau v_4}=\mathrm{i}q_{2}.
	\end{aligned}
\end{align}
Likewise, from \eqref{KP-q3},
\begin{align}
	\begin{aligned}
	q_{3,-}^*
	&=-\frac{\mathrm{i}\,\br_{1,-}^{\dagger}
		(\bM_{-}^{\dagger})^{-1}(\bs_{2,-}^{\st})^{\dagger}}
	{\tau^*v_{4,-}^*}
	=-\frac{\mathrm{i}\,\bs_1^{\st}
		(\bM_{-}^{\dagger})^{-1}\bK^{-1}\br_2}
	{\tau^*v_{4,-}^*}\\
	&=\frac{\mathrm{i}\,\bs_1^{\st}\bM^{-1}\bK^{-1}\br_2}
	{\tau^*v_4v_{4,-}^*}
	=\mathrm{i}\,\tau \bs_1^{\st}\bM^{-1}\bK^{-1}\br_2=-\mathrm{i}q_{4}.
\end{aligned}
\end{align}
We complete the proof.
\end{proof}

The KP-type Cauchy matrix soliton solution and multiple-pole solution
to the nonlocal discrete MTM \eqref{non-DMTM} are summarized in the following theorem. 
\begin{theorem}\label{Theo-KP-solu-non}
Let the constant parameter $\tau$ obey the relation $\tau^2=|\bK|/|\bK^*|$. The functions 
\begin{subequations}
\begin{align}
& q_{1}=-\tau\bs^{\st}_{2}(\bM_{1}+\bM_{2})^{-1}\bK^{-1}\br_{1},\\
& q_{3}=\frac{\mathrm{i}\bs^{\st}_{2}(\bM_{1}+\bM_{2})^{-1}\br_{1}}{\tau(1-\bs^{\st}_{2}(\bM_{1}+\bM_{2})^{-1}(\bs_{2,-}^{\st})^{\dagger})},
\end{align}
\end{subequations}
provide i) the KP-type Cauchy matrix $N$-soliton solution for the nonlocal discrete MTM \eqref{non-DMTM} with
\begin{subequations}
\label{KP-nonlocal-msr-def}
\begin{align}
\label{KP-non-k-def}
&\bK=\mathrm{diag}(k_{1},k_{2},\cdots,k_{N}),\quad k_{i}\in\mathbb{C},\quad i=1,2,\cdots,N,\\
\label{KP-non-sr1-def}
&\bs^{\st}_{2}=\left( \frac{\rho_{2,-}^{*}(k_{1})}{k_{1}^{*}},\frac{\rho_{2,-}^{*}(k_{2})}{k_{2}^{*}},\cdots,\frac{\rho_{2,-}^{*}(k_{N})}{k_{N}^{*}}\right) ,\quad
\br_{1}=(\rho_{1}(k_{1}),\rho_{1}(k_{2}),\cdots,\rho_{1}(k_{N}))^{\st},\\
\label{KP-non-m1-def}
&\bM_{1}=(M_{1,ij})_{N\times N},\quad M_{1,ij}=\frac{\rho_{1}(k_{i})\rho_{1,-}^{*}(k_{j})}{k_{i}-k_{j}^{*}},\\
\label{KP-non-m2-def}
&\bM_{2}=(M_{2,ij})_{N\times N},\quad M_{2,ij}=\frac{\rho_{2}(k_{i})\rho_{2,-}^{*}(k_{j})}{(k_{i}-k_{j}^{*})k_{j}^{*}},
\end{align}
\end{subequations}
and ii) the KP-type Cauchy matrix $N$-pole solution for the nonlocal discrete MTM \eqref{non-DMTM} with
\begin{subequations}
\begin{align}
&\bK=\bJ[k,1],\quad\br_{1}=\bF_{k}[\rho_{1}(k)]\bar{\be}_{N},\quad\bs^{\st}_{2}=\bar{\be}_{N}^{\st}\bF^{\dagger}_{k}[\rho_{2,-}(k)](\bK^{\dagger})^{-1},\\
&\bM_{1}=\bF_{k}[\rho_{1}(k)]\bar{\bG}\bF^{\dagger}_{k}[\rho_{1,-1}(k)],\quad \bM_{2}=\bF_{k}[\rho_{2}(k)]\bar{\bG}
\bF^{\dagger}_{k}[\rho_{2,-}(k)](\bK^{\dagger})^{-1},\\
&\bar{\bG}=(\bar{G}_{i,j})_{N\times N},\quad \bar{G}_{i,j}=\frac{(-1)^{i-1}\mathrm{C}^{i-1}_{i+j-2}}{(k-k^{*})^{i+j-1}},
\end{align}
\end{subequations}
where the discrete plane-wave factors $\rho_{\iota}(z)$ are defined as in equation \eqref{plane-rho-def}, and 
the matrix functions $\bJ,~\bF$ are defined in \eqref{J-def} and \eqref{FH-def}, respectively.
\end{theorem}

\begin{remark}

With the constraint $(q_{2},q_{4})=(-\mathrm{i}q_{1,-}^*,\mathrm{i}q_{3,-}^*)$, 
the semi-discrete four-component MTM \eqref{sdMTM} and the four-component MTM \eqref{MTM}
reduce to the nonlocal semi-discrete MTM
\begin{subequations}
\label{sdMTM-nonlocal}
	\begin{align}
		&\mathrm{i}q_{3,x}-2q_{1}+2\mathrm{i}q_{3}q_{1}q_{1,-}^*=0,\\
		&(q_{3}+\mathrm{i}p\wt{q}_{1})^2-(\wt{q}_{3}-\mathrm{i}pq_{1})^2+\mathrm{i}(q_{1}q_{3}+\wt{q}_{1}\wt{q}_{3})
[q_{3,-}^*(q_{3}+\mathrm{i}p\wt{q}_{1})-\wt{q}_{3,-}^*(\wt{q}_{3}-\mathrm{i}pq_{1})]=0.
	\end{align}
\end{subequations}
and the nonlocal MTM
\begin{subequations}
\label{MTM-nonlocal}
\begin{align}
&\mathrm{i}q_{3,x}-2q_{1}+2\mathrm{i}q_{1}q^*_{1,-}q_{3}=0,\\
&\mathrm{i}q_{1,t}+2q_{3}+2\mathrm{i}q_{1}q_{3}q^*_{3,-}=0,
\end{align}
\end{subequations}
respectively. Theorems \ref{Theo-AKNS-solu-non} and \ref{Theo-KP-solu-non} provide the AKNS-type Cauchy matrix solutions and 
KP-type Cauchy matrix solutions for these reduced systems, respectively, 
upon replacing discrete plane-wave factors \eqref{plane-def} with semi-discrete plane-wave factors \eqref{semi-plane-def}, as well as with continuous exponential functions
\eqref{con-plane-def}.
\end{remark}

\section{Conclusions}

In this paper, we develop the Cauchy matrix scheme to construct a discrete four-component MTM. The construction starts from the DES \eqref{DES}, 
which incorporates a Sylvester equation and two shift relations that determine the dispersion relations. Master functions $\bS^{(i,j)}$
are subsequently introduced, and the associated recurrence and shift relations are derived under the constraint \eqref{kc-cl}. 
Focusing on the recurrence and shift relations among $\bu=\bS^{(0,0)},~\bv=\bI-\bS^{(-1,0)},~
\bw=\bI+\bS^{(0,-1)}$ and $\bh=\bS^{(-1,-1)}$, we eventually obtain a discrete four-component MTM whose dependent variables are defined by \eqref{q1q2q3q4}. 
From the component-wise expansion expressions \eqref{com-wi-1}-\eqref{com-wi-4}, 
we derive the relation between the dependent variables of the discrete four-component MTM system and those of the discrete Mikhailov model.
By taking two successive continuum limits, the discrete four-component MTM reduces first to a semi-discrete four-component MTM and then to the continuous four-component MTM.
With the constraint \eqref{kc-cl}, we further construct AKNS-type and KP-type solutions, including explicit soliton solutions and multiple-pole solutions. 
Imposing the local reduction $(q_{2}, q_{4})=(q^*_{1}, q^*_{3})$ and the nonlocal reduction $(q_{2}, q_{4})=(-\mathrm{i}q^*_{1,-}, \mathrm{i}q^*_{3,-})$,
the discrete four-component MTM reduces into the discrete MTM and a reverse-$(n,m)$ discrete MTM, respectively. 
For these two reduced systems, we prove that the former admits AKNS-type solutions, while the latter possesses both AKNS-type and KP-type solutions.

We conclude the paper with the following remarks. First, this work only constructs explicit soliton solutions and multiple-pole solutions 
for the derived discrete four/two-component MTM, while a detailed analysis of the lattice dynamics of these solutions is not carried out. 
A systematic investigation of their velocities, phase shifts, singularity conditions and collision dynamics would help elucidate how 
discretization and nonlocality alter the dynamics of the continuous MTM.
In addition, Tsuchida proposed a discrete counterpart of the four-component MTM via constructing its Lax pair representation in \cite{T2024}. 
It remains an interesting open problem to construct the Lax pair for the discrete four-component MTM presented in this paper.
Lastly, bilinearization, Casoratian-type solutions and algebraic solutions for the resulting systems are topics worthy of consideration in future research.

\vskip 20pt
\section*{Acknowledgments}
This project is supported by National Natural Science Foundation of China (No. 12071432) 
and Natural Science Foundation of Zhejiang Province (No. LZ24A010007).

\vskip 20pt
\section*{Data Availability Statement}
Data sharing not applicable to this article as no datasets were generated or
analyzed during the current study.

\vskip 20pt
\section*{Conflict of interest}
There are no conflicts of interest to declare.

\end{document}